\documentclass[journal]{IEEEtran}

\usepackage{amsmath,amssymb,amsthm}
\usepackage{algorithm,algorithmicx,algpseudocode}
\usepackage{cite}
\usepackage{graphicx}
\usepackage{array}
\usepackage{balance}
\usepackage{color,soul}
\usepackage{lipsum}
\usepackage{bm}
\usepackage{amsmath, flushend}
\usepackage[caption=false,font=footnotesize]{subfig}

\PassOptionsToPackage{bookmarks=false}{hyperref}
\usepackage[hidelinks]{hyperref}

\theoremstyle{plain}

\theoremstyle{definition}

\newcommand{\bmI}{\mathbf I}

\newcommand{\bmh}{\mathbf h}

\usepackage{dblfloatfix}
\usepackage{cuted}

 \usepackage{tikz}
\usetikzlibrary{arrows.meta,positioning,fit,calc,decorations.pathmorphing}
\usepackage{caption}

\usepackage[normalem]{ulem} 
\usepackage[user]{zref}

\newcommand{\sot}[1]{} 

\newcounter{revc}
\makeatletter
\zref@newprop{revcontent}{} \zref@addprop{main}{revcontent}
\zref@newprop{revsec}{} \zref@addprop{main}{revsec}
\zref@newprop{revpage}{} \zref@addprop{main}{revpage}
\makeatother

\newcommand{\revi}[2]{\zref@setcurrent{revsec}{\thesection}%
\zref@setcurrent{revpage}{\thepage}%
\zref@setcurrent{revcontent}{#2}%
\refstepcounter{revc}%
\label{#1}%
\zlabel{#1}%
\textcolor{blue}{#2}}

\newcommand{\revinu}[2]{\zref@setcurrent{revsec}{\thesection}%
\zref@setcurrent{revcontent}{#2}%
\refstepcounter{revc}%
\zlabel{#1}\label{#1}#2 }

\newcommand{\revr}[2]{\zref@setcurrent{revsec}{\thesection}%
\zref@setcurrent{revcontent}{#2}%
\refstepcounter{revc}%
\zlabel{#1}\label{#1}\sot{#2}}

\begin{document}
\title{STAR Beyond Diagonal RISs with Amplification:\\ Modeling and Optimization}
\author{Chandan~Kumar~Sheemar,~\IEEEmembership{Member,~IEEE}, Giovanni~Iacovelli,~\IEEEmembership{Member,~IEEE}, Wali Ullah Khan,~\IEEEmembership{Member,~IEEE}, \\George C. Alexandropoulos,~\IEEEmembership{Senior~Member,~IEEE}, Stefano Tomasin,~\IEEEmembership{Senior~Member,~IEEE}, \\and Symeon Chatzinotas,~\IEEEmembership{Fellow,~IEEE}  
 \thanks{C. K. Sheemar, G. Iacovelli, W. U. Khan, and  Symeon Chatzinotas are with the SnT, University of Luxembourg, 1855 Luxembourg City, Luxembourg (emails:\{chandankumar.sheemar, giovanni.iacovelli, wali.ullah.khan, symeon.chatzinotas\}@uni.lu).
 G. C. Alexandropoulos is with the Department of Informatics and Telecommunications, National and Kapodistrian University of Athens, 16122 Athens, Greece and with the Department of Electrical and Computer Engineering, University of Illinois Chicago, IL 60601, USA \{email: alexandg@di.uoa.gr\}. S. Tomasin is with the University of Padua, Italy, \{email: tomasin@dei.unipd.it\}} }
 \maketitle

\begin{abstract}
 This paper develops a physically consistent signal model with hardware constraints for a simultaneous transmitting and reflecting beyond-diagonal RIS (STAR BD-RIS) endowed with per-element amplification and lossless power splitting. We explicitly decouple (i) amplification via a diagonal gain matrix, (ii) element-wise reflection/transmission splitting, and (iii) passive beyond-diagonal coupling on each branch, while enforcing practical feasibility through per-element emission caps and an aggregate RIS power budget under the operating covariance. Building on this model, we cast downlink sum-rate maximization as an equivalent weighted minimum mean-square error (WMMSE) problem and propose an alternating optimization framework with provable monotonic descent. The method admits closed-form updates for MMSE combiners and weights, waterfilling-like beamformer updates via a single dual variable, a per-element amplification update that satisfies emission constraints, and a STAR power-splitting update based on cyclic coordinate descent with a global acceptance test. For the beyond-diagonal coupling matrices, we derive Riemannian gradient steps on the complex Stiefel manifold with QR/polar retraction method, preserving passivity at every iterate. Furthermore, the proposed approach decouples the optimization of the reflective and transmissive responses of the BD-RIS, enabling efficient distributed implementation. Numerical results demonstrate substantial sum-rate gains compared to the conventional passive BD-RIS. 
\end{abstract}
 
 \section{Introduction}
\IEEEpeerreviewmaketitle
 
\IEEEPARstart{R}{econfigurable} intelligent surfaces (RISs) have emerged as a transformative technology capable of fundamentally reshaping wireless propagation environments \cite{huang2019reconfigurable,di2019smart}. An RIS comprises a large array of meta elements whose electromagnetic responses can be dynamically tuned via low-power control circuitry. By intelligently adjusting these responses, an RIS can manipulate incident radio waves to achieve desired scattering behaviors such as reflection, refraction, or beam focusing, thereby enabling the surface to constructively reconfigure the effective wireless channel \cite{jian2022reconfigurable}. This capability opens new possibilities for extending coverage, mitigating interference, and enhancing signal strength in complex scenarios such as dense urban deployments, millimeter-wave networks, and indoor non-line-of-sight environments. Unlike conventional relays or active antenna arrays, RISs operate without generating new signals or performing baseband processing. Their near-passive nature allows them to achieve significant performance improvements with extremely low power consumption, making them an attractive and sustainable solution for future 6G and beyond communication systems \cite{10596064,fotock2023energy,khan2024reconfigurable}. 
 
The simplest and most widely studied RIS architecture is the diagonal RIS (D-RIS), where each of the $N$ surface elements operates independently and applies a controllable complex reflection coefficient to the incident field \cite{Tsinghua_RIS_Tutorial,lin2020reconfigurable}. In this case, the overall reflection operation is represented by a diagonal matrix $\boldsymbol{\Phi}=\operatorname{diag}(e^{j\theta_1},\ldots,e^{j\theta_N})$, where $\theta_n$ denotes the tunable phase shift applied by element $n$ \cite{liu2021reconfigurable}. This configuration enables coherent signal combining at the intended receiver through coordinated phase alignment of the reflected wavefront. While diagonal RISs are conceptually simple and easy to implement, their element-wise independence inherently limits the degrees of freedom in wave manipulation \cite{hou2020reconfigurable,basar2024reconfigurable}. In particular, they cannot exploit mutual coupling or inter-element correlations that arise in realistic electromagnetic environments, restricting the achievable control over the spatial and polarization domains of the scattered field.
 
To overcome these limitations, the concept of a beyond-diagonal RIS (BD-RIS) has recently been introduced. In BD-RISs, the reflection matrix $\boldsymbol{\Phi}$ is no longer diagonal but instead structured to include off-diagonal entries that model electromagnetic coupling between different elements \cite{khan2025survey,fang2023low}. This allows each element’s excitation to influence multiple ports of the surface, effectively realizing a linear network with distributed reflection and transmission paths \cite{nerini2023closed,li2023beyond}. Such coupling can be achieved through embedded analog circuitry, e.g., impedance networks, metamaterial interconnections, or hybrid couplers, which enable more flexible and energy-efficient wave transformations \cite{nerini2024beyond}. Consequently, BD-RIS architectures can jointly optimize amplitude, phase, and coupling patterns across the surface, achieving superior beamforming precision and richer spatial control compared to D-RISs \cite{li2024beyond,zhou2023optimizing,wu2025beyond}. 
 
The inherent limitation of conventional reflective-only RISs has recently motivated the development of the simultaneous transmitting and reflecting RIS (STAR-RIS) paradigm \cite{liu2022star}, which enables each surface element to concurrently control both the reflected and transmitted components of the incident wave \cite{ahmed2023survey,mu2021simultaneously}. This structure falls in the category of dual-functional RISs. By dynamically adjusting the amplitude and phase of these two components, a STAR-RIS can serve users located on both sides of the surface, thereby realizing a full-space, programmable wireless environment \cite{liu2022star}. 

\subsection{State-of-the-Art and Motivation} 

The following section presents an overview of the existing literature, highlighting key studies that have advanced the development of STAR-RIS technologies toward a sustainable wireless ecosystem.

\subsubsection{On the STAR D-RISs}
In \cite{mu2021simultaneously}, the authors analyzed the potential of STAR RIS for improved coverage by splitting incident signals into reflected and transmitted components. Three practical operation protocols are proposed: energy splitting, mode switching, and time switching, along with a joint beamforming optimization framework. In \cite{xu2021star}, the authors presented a general hardware model for STAR-RISs and channel modeling for the case of near-field and far-field scenarios. In \cite{wu2021channel}, efficient uplink channel estimation schemes are proposed for STAR-RIS–assisted two-user systems under the time-switching and energy-splitting protocols. In \cite{niu2021weighted}, the authors studied the STAR-RIS-assisted MIMO system, where the weighted sum rate is maximized under the energy splitting scheme. A sub-optimal block coordinate descent (BCD) algorithm is proposed, combining Lagrange dual optimization for transmit–reflect coefficient design. In \cite{fang2022energy}, 
a STAR-RIS–assisted MIMO non-orthogonal multiple-access (NOMA) system for energy-efficient resource allocation is studied. A novel algorithm to jointly optimize the base station (BS) beamforming and the STAR-RIS phase shifts is presented. In \cite{li2022enhancing}, authors studied the secrecy performance of STAR-RIS–assisted NOMA networks considering residual hardware impairments over Nakagami$-m$ fading channels. Closed-form and asymptotic expressions for the secrecy outage probability to characterize system performance at high signal-to-noise (SNR) are also derived. In \cite{zhang2022joint}, a STAR-RIS–assisted UAV communication system is investigated, and a novel algorithm aimed at maximizing the sum rate of the users through joint optimization of the STAR-RIS beamforming, UAV trajectory, and power allocation is presented.

Active STAR-RISs, in which the surface elements are equipped with low-power amplification \cite{ma2023optimization}, have also been widely investigated due to their potential to compensate for severe path losses, enhance signal coverage, and enable simultaneous transmission and reflection with controllable power gain, thereby extending the functionality and efficiency of conventional passive STAR-RIS architectures \cite{ma2023active}. They fall into the category of multifunctional RIS\footnote{A passive STAR-RIS belongs to the class of dual-functional RISs, as it simultaneously supports reflection and transmission.
In contrast, multi-functional RISs extend this concept by enabling more than two operational modes.
Accordingly, an active STAR-RIS, which incorporates amplification, is classified as a multi-functional RIS.}\cite{ni2024single}. In the study \cite{zhang2024joint}, an active STAR-RIS–assisted full-duplex integrated sensing and communication (ISAC) system is analyzed and a novel alternating optimization-based method is proposed to maximize the communication sum rate while satisfying the radar SINR, hardware, and power constraints. In \cite{papazafeiropoulos2024two}, the authors propose Active STAR RIS for mMIMO systems by integrating active amplification to overcome the double-fading effect. A novel two-time scale optimization framework to jointly optimize the amplitudes, phases, and gains under correlated fading is proposed. In \cite{faramarzi2025energy}, the authors presented a novel method based on alternating optimization combined with reinforcement learning to optimize the performance of the active STAR RIS-assisted simultaneous wireless information and power transfer (SWIPT) system. 

\subsubsection{On the STAR BD-RISs} 
We remark that the literature on STAR BD-RIS is limited to the passive case \cite{li2022beyond,li2023dynamic,ming2025hybrid,mahmood2024enhancing}.
It has been shown in \cite{li2022beyond} that the conventional STAR-RIS represents a special case of a block-diagonal BD-RIS, which can be physically realized using a group-connected reconfigurable impedance network, where each pair of ports is interconnected. In \cite{li2023dynamic}, a dynamically group-connected BD-RIS architecture is proposed, where RIS elements are adaptively partitioned into subsets based on channel state information (CSI), forming a permuted block-diagonal scattering matrix. This adaptive design notably outperforms fixed group-connected architectures in multi-user MISO systems. Furthermore, \cite{ming2025hybrid} introduces a hybrid STAR BD-RIS capable of independent and tunable beam steering for both reflection and transmission within a shared aperture. The design employs phase-reconfigurable antenna arrays interconnected by tunable two-port power splitters, with experimental validation of a 4×4 prototype demonstrating precise dual-beam control and bridging electromagnetic modeling with practical hardware realization. Finally, in \cite{mahmood2024enhancing} the potential of STAR BD-RIS
for joint indoor and outdoor communications in the THz-band have been investigated.

Although recent studies have investigated passive STAR BD-RIS architectures, it is noteworthy that the research on passive STAR BD-RISs still remains limited, while the case of active STAR BD-RIS remains unexplored. Existing works on passive STAR BD-RISs have demonstrated promising results, such as adaptive inter-element coupling, dynamic grouping based on CSI, and independent beam steering across reflection and transmission paths, highlighting their potential for fine-grained spatial control and full-space coverage. However, these designs remain constrained by their passive nature, which inherently limits signal amplification and coverage extension, particularly in large-scale or severe path-loss communication scenarios. This gap motivates the introduction of the active STAR BD-RIS, which integrates low-power amplification into the hybrid transmitting and reflecting BD-RIS framework. By enabling joint control over amplitude, phase, and inter-element coupling across both reflective and transmissive domains, the active STAR BD-RIS is envisioned to overcome the inherent double-fading limitation of passive surfaces, significantly improve energy efficiency, and realize adaptive, power-reconfigurable full-space communications. This advancement not only surpasses existing passive STAR BD-RIS designs but also establishes a new class of multifunctional BD-RIS architectures capable of simultaneously shaping, amplifying, and redistributing electromagnetic energy in a fully programmable manner.

\subsection{Main Contributions}
We consider a multi-user multiple-input single-output (MISO) communication system assisted with an active STAR BD-RIS. The BD-RIS is assumed to be equipped with very low-power amplifiers per cell, with each cell containing one transmitting and one reflecting element \cite{li2022beyond}. The conventional modeling and design constraints established for passive BD-RIS architectures cannot be directly applied to the active STAR BD-RIS paradigm proposed in this work \cite{li2022beyond,li2023dynamic,ming2025hybrid,mahmood2024enhancing}. In passive STAR BD-RIS systems, the surface merely redistributes the impinging electromagnetic energy under a strict power-conservation law enforced through right-unitary coupling matrices, guaranteeing lossless operation for any excitation. However, the introduction of active components alters this behavior: conservation is no longer unconditional but instead governed by hardware-aware feasibility constraints that depend on the operating covariance of the surface. In this context, we first develop an accurate and physically consistent system model that explicitly accounts for per-element amplification, power splitting between transmitting and reflective modes, and beyond-diagonal coupling within both reflective and transmissive branches. This formulation leads to a new set of active feasibility constraints that jointly capture per-element emission limits, total radiated power budgets, and lossless energy splitting conditions. 

We then consider the sum-rate maximization problem under the newly introduced active STAR BD-RIS feasibility constraints, which intricately couple amplification, power splitting, and beyond-diagonal coupling effects. To efficiently address this non-convex problem, we reformulate it within the weighted minimum mean-square error (WMMSE) framework, enabling an alternating optimization approach that ensures monotonic convergence. Each system variable is updated through a specialized and theoretically grounded method: the MMSE combiners and weights are derived in closed form; the BS's beamformers are optimized via a waterfilling-type update governed by a single dual variable; the per-element amplification factors are refined through a gradient-based projection method that guarantees compliance with both per-element and total emission power constraints; and the power-splitting ratios are optimized using a cyclic coordinate descent scheme equipped with a global acceptance test to ensure monotonic improvement. Finally, the beyond-diagonal coupling matrices, which reside on the complex Stiefel manifold, are updated through Riemannian gradient descent with QR/polar retraction method. Furthermore, the proposed approach decouples the optimization of the reflective and transmissive responses of the BD-RIS, thereby enabling an efficient distributed implementation. Numerical results demonstrate that the proposed active STAR BD-RIS significantly outperforms the passive STAR BD-RIS-assisted system.

\emph{Paper Organization:} The rest of the paper is organized as follows. Section \ref{section_2} presents the novel constraints set for active STAR RIS, presents the system model, and formulates the optimization problem. Section \ref{section_3} proposes a novel optimization framework to solve the problem. Finally, Sections \ref{section_4} and \ref{section_5} present the simulation results and conclusions, respectively.

\emph{Notations:} In this paper, we adopt a consistent set of notations: Scalars are denoted by lowercase or uppercase letters, while vectors and matrices are represented by bold lowercase and bold uppercase letters, respectively. The transpose, Hermitian transpose, and inverse of a matrix $\mathbf{X}$ are denoted by $\mathbf{X}^\mathrm{T}$, $\mathbf{X}^\mathrm{H}$, and $\mathbf{X}^{-1}$, respectively. Sets are indicated either by calligraphic letters (e.g., $\mathcal{X}$) or $\{\cdots\}$.  

\section{System Model} \label{section_2} 
\subsection{On the STAR BD-RIS Operation with Amplification} \label{subsec:multi_functional_bdris}
Let the BS and the BD-RIS be equipped with $M$ transmit antennas and $N$ cells with each cell containing one transmit and one reflective element, respectively. We begin from the conventional passive BD-RIS, which redistributes only the received electromagnetic energy across its $N$ ports. Let the incident field at the surface be:
\begin{equation}
\mathbf r=\mathbf G\mathbf x, \qquad \boldsymbol{\Sigma}_r \triangleq \mathbb E[\mathbf r\mathbf r^{H}],
\end{equation}
where $\mathbf G\in\mathbb C^{N\times M}$ is the BS--RIS channel and $\mathbf x\in\mathbb C^{M}$ is the BS transmit signal. A passive STAR BD-RIS splits and mixes this incident wavefront into reflective and transmissive branches through fully connected coupling networks $\boldsymbol{\Phi}_{\mathrm R},\boldsymbol{\Phi}_{\mathrm T}\in\mathbb C^{N\times N}$, yielding:
\begin{equation}
\mathbf y_{\mathrm R}^{(\text{pass})}=\boldsymbol{\Phi}_{\mathrm R}\mathbf r,\qquad
\mathbf y_{\mathrm T}^{(\text{pass})}=\boldsymbol{\Phi}_{\mathrm T}\mathbf r.
\end{equation}
Because the network is lossless, the overall transformation is right-unitary:
\begin{equation}
\label{eq:passive_unitarity}
\boldsymbol{\Phi}_{\mathrm R}^{H}\boldsymbol{\Phi}_{\mathrm R}
+\boldsymbol{\Phi}_{\mathrm T}^{H}\boldsymbol{\Phi}_{\mathrm T}=\mathbf I_N,
\end{equation}
which guarantees conservation of power for any excitation.
Hence, $\boldsymbol{\Phi}_{\mathrm R}$ and $\boldsymbol{\Phi}_{\mathrm T}$ act as lossless energy mixers in the passive case.

To enhance link budgets and extend coverage, we upgrade the surface to actively amplify the impinging field before splitting and coupling. This fundamentally changes modeling: conservation is no longer unconditional; instead, operation must satisfy feasibility constraints tied to the operating covariance and hardware power budgets. We therefore separate the signal flow into three modules:
(i) per-element amplification, (ii) lossless per-element splitting, and (iii) passive beyond-diagonal coupling per branch.

We assume that each cell, containing the transmitting and reflecting elements, is equipped with independently integrated low-cost amplifiers to amplify the signal. Let the per-element amplitude gains be $\beta_i=\sqrt{a_i}\ge 1$ and define:
\begin{equation}
\mathbf A=\operatorname{diag}(\beta_1,\ldots,\beta_N).
\end{equation}
Assuming lossless per-element split, the amplified field is divided between reflection and transmission with the energy splitting matrices $\mathbf{E}_R$ and $\mathbf{E}_T$, given as:
\begin{equation}
\mathbf E_{\mathrm R}=\operatorname{diag}(\varsigma_1,\ldots,\varsigma_N),
\end{equation}
\begin{equation}
\mathbf{E}_{\mathrm T}=\operatorname{diag}\!\big(\sqrt{1-\varsigma_1^2},\ldots,\sqrt{1-\varsigma_N^2}\big),
\end{equation}
with $\varsigma_i\in[0,1]$, which enforce elementwise energy conservation at the splitter for the amplified power as:
\begin{equation}
\label{eq:split_cons}
\mathbf E_{\mathrm R}^{H}\mathbf E_{\mathrm R}+\mathbf E_{\mathrm T}^{H}\mathbf E_{\mathrm T}=\mathbf I_N.
\end{equation}
After amplification and energy splitting, each branch decouples and the signal is spatially mixed by a beyond-diagonal network. To avoid any additional power creation beyond $\mathbf A$, we impose per-branch right-unitarity condition to conserve the splitted energy as:
\begin{equation}
\label{eq:per_branch_unitary}
\boldsymbol{\Phi}_{\mathrm R}^{H}\boldsymbol{\Phi}_{\mathrm R}=\mathbf I_N,\qquad
\boldsymbol{\Phi}_{\mathrm T}^{H}\boldsymbol{\Phi}_{\mathrm T}=\mathbf I_N.
\end{equation}
Thus, all gain is attributed to $\mathbf A$, the split is lossless via $(\mathbf E_{\mathrm R},\mathbf E_{\mathrm T})$, and each coupling network is passive. 

Note that as amplifiers are active components, consequently, amplification  stage forwards both the signal and the internal noise generated through active components. Let the input to the active core be:
\begin{equation}
\mathbf v=\mathbf r+\mathbf n,\qquad \mathbf n\sim\mathcal{CN}(\mathbf 0,\mathbf N), \qquad
\boldsymbol{\Sigma}_v\triangleq \mathbb E[\mathbf v\mathbf v^{H}],
\end{equation}
where $\mathbf N=\operatorname{diag}(\sigma_1^2,\ldots,\sigma_N^2)$ models the equivalent noise introduced at the RIS ports. The scattered fields are then given as~\footnote{Here we assume a cascade of amplification, power split, and spatial mixing. Other cascades of these operations are possible, but left for future study.}: 
\begin{subequations} \label{eqRIS}   
\begin{align}
\mathbf y_{\mathrm R}&=\boldsymbol{\Phi}_{\mathrm R}\,\mathbf E_{\mathrm R}\,\mathbf A\,\mathbf v, \quad
\mathbf{y}_{\mathrm T} =\boldsymbol{\Phi}_{\mathrm T}\,\mathbf E_{\mathrm T}\,\mathbf A\,\mathbf v,
\end{align}
\end{subequations}
and covariance propagation yields:
\begin{subequations}
\label{eq:active_covs}
\begin{align}
\boldsymbol{\Sigma}_{y,\mathrm R}&=\boldsymbol{\Phi}_{\mathrm R}\,\mathbf E_{\mathrm R}\,\mathbf A\,\boldsymbol{\Sigma}_v\,\mathbf A\,\mathbf E_{\mathrm R}\,\boldsymbol{\Phi}_{\mathrm R}^{H},\\
\boldsymbol{\Sigma}_{y,\mathrm T}&=\boldsymbol{\Phi}_{\mathrm T}\,\mathbf E_{\mathrm T}\,\mathbf A\,\boldsymbol{\Sigma}_v\,\mathbf A\,\mathbf E_{\mathrm T}\,\boldsymbol{\Phi}_{\mathrm T}^{H}.
\end{align}
\end{subequations}
Define the per-element emitted powers (sum of reflective and transmissive emissions) as
\begin{equation}
\label{eq:per_element_power_def_para}
\mathbf P_{\mathrm{out}}
=
\operatorname{diag}\!\Big(
\boldsymbol{\Phi}_{\mathrm R}\mathbf E_{\mathrm R}\mathbf A\,\boldsymbol{\Sigma}_v\,\mathbf A\,\mathbf E_{\mathrm R}\boldsymbol{\Phi}_{\mathrm R}^{H}
+
\boldsymbol{\Phi}_{\mathrm T}\mathbf E_{\mathrm T}\mathbf A\,\boldsymbol{\Sigma}_v\,\mathbf A\,\mathbf E_{\mathrm T}\boldsymbol{\Phi}_{\mathrm T}^{H}
\Big),
\end{equation}
Let $\mathbf P_{\max}=\operatorname{diag}(P_{\max,1},\ldots,P_{\max,N})$ denote per-element amplifier ratings and thermal limits, with $P_{\max,i}$ denoting the maximum power limit of the amplifier in the $i$-th cell, leading to the per-element emission cap:
\begin{equation}
\label{eq:per_element_cap_para}
\begin{aligned}
(C1)\quad \operatorname{diag}\!\Big( &
\boldsymbol{\Phi}_{\mathrm R}\mathbf E_{\mathrm R}\mathbf A\,\boldsymbol{\Sigma}_v\,\mathbf A\,\mathbf E_{\mathrm R}\boldsymbol{\Phi}_{\mathrm R}^{H}
\\& +
\boldsymbol{\Phi}_{\mathrm T}\mathbf E_{\mathrm T}\mathbf A\,\boldsymbol{\Sigma}_v\,\mathbf A\,\mathbf E_{\mathrm T}\boldsymbol{\Phi}_{\mathrm T}^{H}
\Big)
\le
\mathbf P_{\max},
\end{aligned}
\end{equation}
and the aggregate supply/regulatory constraint yields the total emission budget
\begin{equation}
\label{eq:total_cap_para}
\begin{aligned}
(C2) \quad \operatorname{Tr}\!\Big(&
\boldsymbol{\Phi}_{\mathrm R}\mathbf E_{\mathrm R}\mathbf A\,\boldsymbol{\Sigma}_v\,\mathbf A\,\mathbf E_{\mathrm R}\boldsymbol{\Phi}_{\mathrm R}^{H}
\\&+
\boldsymbol{\Phi}_{\mathrm T}\mathbf E_{\mathrm T}\mathbf A\,\boldsymbol{\Sigma}_v\,\mathbf A\,\mathbf E_{\mathrm T}\boldsymbol{\Phi}_{\mathrm T}^{H}
\Big)
\le
P_{\max}.
\end{aligned}
\end{equation}
In parallel, the lossless internal power division between reflection and transmission is imposed, enforcing the following constraints:
\begin{equation}
\label{eq:split_cons_para}
(C3) \quad\mathbf E_{\mathrm R}^{H}\mathbf E_{\mathrm R}
+
\mathbf E_{\mathrm T}^{H}\mathbf E_{\mathrm T}
=
\mathbf I_N , \quad
  (C4) \quad 0\le \varsigma_i \le 1,
\end{equation}
and the beyond-diagonal coupling networks in each branch remain passive in response to the amplified signal, leading to the constraints
\begin{equation}
\label{eq:branch_passive_cons_para}
(C5)\quad \boldsymbol{\Phi}_{\mathrm R}^{H}\boldsymbol{\Phi}_{\mathrm R}
=
\mathbf I_N, \quad 
(C6)\quad\boldsymbol{\Phi}_{\mathrm T}^{H}\boldsymbol{\Phi}_{\mathrm T}
=
\mathbf I_N,
\end{equation}
Collectively, \eqref{eq:per_element_cap_para}–\eqref{eq:branch_passive_cons_para} constitute the active analogue of \eqref{eq:passive_unitarity}: conservation is enforced after amplification and splitting, and relative to the actual operating covariance $\boldsymbol{\Sigma}_v$ through per-element and total power budgets.

\subsection{Signal Model}
\begin{figure}
    \centering
\includegraphics[width=0.9\linewidth]{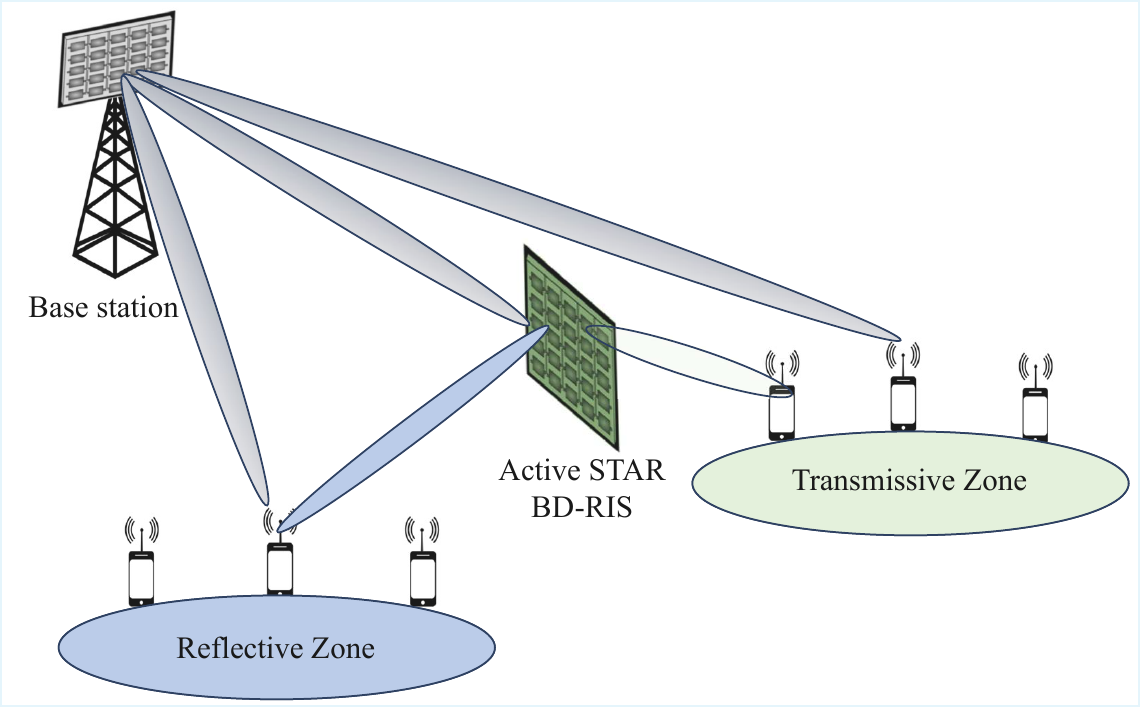}
    \caption{Multi-Functional BD-RIS-assisted communications system with simultaneous transmission, reception, and amplification.}
    \label{scenario}
\end{figure}
We consider a downlink multi-user system aided by a multi-functional BD-RIS, as shown in Figure \ref{scenario}. The BS equipped with $M$ transmit antennas serves $K$ single-antenna users that are spatially separated into the transmissive and reflective zones of the multi-functional BD-RIS having $N$ cells, with each cell having both the transmissive and reflective elements. The user indices in each zone are collected in the following two disjoint sets:
\begin{equation}
    \mathcal{K}_{\mathrm{T}} = \{1,2,\ldots,K_{\mathrm{T}}\}, 
    \quad 
    \mathcal{K}_{\mathrm{R}} = \{K_{\mathrm{T}}+1, K_{\mathrm{T}}+2, \ldots, K\},
\end{equation}
where $K_{\mathrm{T}}$ and $K_{\mathrm{R}}$ denote the numbers of transmissive-zone and reflective-zone users, respectively, with:
\begin{equation}
    K_{\mathrm{T}} + K_{\mathrm{R}} = K, 
    \qquad 
    \mathcal{K}_{\mathrm{T}} \cap \mathcal{K}_{\mathrm{R}} = \emptyset.
\end{equation}
Let $s_{k_t} \sim \mathcal{CN}(0,1)$ and $s_{k_r} \sim \mathcal{CN}(0,1)$ denote the information-bearing symbol intended for a transmissive-zone user $k_t \in \mathcal{K}_{\mathrm{T}}$ and reflective-zone user $k_r \in \mathcal{K}_{\mathrm{R}}$, respectively. To transmit these symbols, the BS employs linear precoding, where $\mathbf{w}_{k_t} \in \mathbb{C}^{M \times 1}$ is the beamforming vector associated with user $k_t$, and $\mathbf{w}_{k_r} \in \mathbb{C}^{M \times 1}$ is the beamforming vector associated with user $k_r$. The transmitted signal from the BS is thus
\begin{equation}
    \mathbf{x} = \sum_{k_t \in \mathcal{K}_{\mathrm{T}}} \mathbf{w}_{k_t} s_{k_t} + \sum_{k_r \in \mathcal{K}_{\mathrm{R}}} \mathbf{w}_{k_r} s_{k_r}.
\end{equation}
Let $\mathcal{K}\triangleq \mathcal{K}_{\mathrm{T}}\cup\mathcal{K}_{\mathrm{R}}=\{1,\ldots,K\}$. For a reflective-zone user $k_r\in\mathcal{K}_{\mathrm{R}}$, its received signal is denoted as $y_{k_r}$,which is given as:
 \begin{equation}
\label{eq:ykr_separated}
\begin{aligned}
    y_{k_r}
    &= 
    \Big(\mathbf{h}_{k_r}^H 
    + \mathbf{g}_{k_r}^H 
      \boldsymbol{\Phi}_{\mathrm{R}}
      \mathbf{E}_{\mathrm{R}}
      \mathbf{A}
      \mathbf{G}\Big)
    \mathbf{w}_{k_r}s_{k_r}
    \\
    &\quad+\;
    \sum_{j\in\mathcal{K}\setminus\{k_r\}}
    \Big(\mathbf{h}_{k_r}^H 
    + \mathbf{g}_{k_r}^H 
      \boldsymbol{\Phi}_{\mathrm{R}}
      \mathbf{E}_{\mathrm{R}}
      \mathbf{A}
      \mathbf{G}\Big)
    \mathbf{w}_{j}s_{j}
    \\
    &\quad+\;
    \mathbf{g}_{k_r}^H 
    \boldsymbol{\Phi}_{\mathrm{R}}
    \mathbf{E}_{\mathrm{R}}
    \mathbf{A}\,\mathbf{n}
    + n_{k_r},
\end{aligned}
\end{equation}
where $\mathbf{h}_{k_r}\in\mathbb{C}^{M\times 1}$ is the BS–user channel, $\mathbf{g}_{k_r}\in\mathbb{C}^{N\times 1}$ is the RIS–user channel (reflective side),
, and $n_{k_r} \sim \mathcal{CN}(0, \sigma_{k_r}^2)$ is the receiver noise. For a transmissive-zone user $k_t\in\mathcal{K}_{\mathrm{T}}$, its received signal denoted as $y_{k_t}$ is given by:
\begin{equation}
\label{eq:ykt_separated}
\begin{aligned}
    y_{k_t}
    &=
    \Big(\mathbf{h}_{k_t}^H
    + \mathbf{g}_{k_t}^H
      \boldsymbol{\Phi}_{\mathrm{T}}
      \mathbf{E}_{\mathrm{T}}
      \mathbf{A}
      \mathbf{G}\Big)
    \mathbf{w}_{k_t}s_{k_t}
    \\
    &\quad+\;
    \sum_{j\in\mathcal{K}\setminus\{k_t\}}
    \Big(\mathbf{h}_{k_t}^H
    + \mathbf{g}_{k_t}^H
      \boldsymbol{\Phi}_{\mathrm{T}}
      \mathbf{E}_{\mathrm{T}}
      \mathbf{A}
      \mathbf{G}\Big)
    \mathbf{w}_{j}s_{j}
    \\
    &\quad+\
    \mathbf{g}_{k_t}^H
    \boldsymbol{\Phi}_{\mathrm{T}}
    \mathbf{E}_{\mathrm{T}}
    \mathbf{A}\,\mathbf{n}
    + n_{k_t},
\end{aligned}
\end{equation}
where $\mathbf{h}_{k_t} \in \mathbb{C}^{M \times 1}$ is the direct channel from the BS to user $k_t \in \mathcal{K}_T$, $\mathbf{g}_{k_t} \in \mathbb{C}^{N \times 1}$ is the channel from the BD-RIS transmissive side to user $k_t$, and $n_{k_t} \sim \mathcal{CN}(0,\sigma_{k_t}^2)$ is the receiver noise. Let  $\gamma_{k_r}$ and $\gamma_{k_t}$ denote the signal-to-interference-plus-noise ratio (SINR) at the reflective user $k_r$ and tranmissive user $k_t$, given in equations \eqref{eq:SINR_reflective} and \eqref{eq:SINR_transmissive}, respectively, at the top of the next page.

 \begin{figure*}[!t]
\begin{equation}
\label{eq:SINR_reflective}
\gamma_{k_r}
= \frac{\Big|\big(\mathbf{h}_{k_r}^H+\mathbf{g}_{k_r}^H \boldsymbol{\Phi}_{\mathrm{R}} \mathbf{E}_{\mathrm{R}} \mathbf{A} \mathbf{G}\big)\mathbf{w}_{k_r}\Big|^2}
{\sum\limits_{j\in\mathcal{K}\setminus\{k_r\}}
 \Big|\big(\mathbf{h}_{k_r}^H+\mathbf{g}_{k_r}^H \boldsymbol{\Phi}_{\mathrm{R}} \mathbf{E}_{\mathrm{R}} \mathbf{A} \mathbf{G}\big)\mathbf{w}_{j}\Big|^2
 \;+\; \sigma_{k_r}^2
 \;+\; \mathbf{g}_{k_r}^H \boldsymbol{\Phi}_{\mathrm{R}} \mathbf{E}_{\mathrm{R}} \mathbf{A}\,\mathbf{N}\,\mathbf{A}\mathbf{E}_{\mathrm{R}} \boldsymbol{\Phi}_{\mathrm{R}}^{H} \mathbf{g}_{k_r} }.
\end{equation}
\begin{equation}
\label{eq:SINR_transmissive}
\gamma_{k_t}
= \frac{\Big|\big(\mathbf{h}_{k_t}^H+\mathbf{g}_{k_t}^H \boldsymbol{\Phi}_{\mathrm{T}} \mathbf{E}_{\mathrm{T}} \mathbf{A} \mathbf{G}\big)\mathbf{w}_{k_t}\Big|^2}
{\sum\limits_{j\in\mathcal{K}\setminus\{k_t\}}
 \Big|\big(\mathbf{h}_{k_t}^H+\mathbf{g}_{k_t}^H \boldsymbol{\Phi}_{\mathrm{T}} \mathbf{E}_{\mathrm{T}} \mathbf{A} \mathbf{G}\big)\mathbf{w}_{j}\Big|^2
 \;+\; \sigma_{k_t}^2
 \;+\; \mathbf{g}_{k_t}^H \boldsymbol{\Phi}_{\mathrm{T}} \mathbf{E}_{\mathrm{T}} \mathbf{A}\,\mathbf{N}\,\mathbf{A}\mathbf{E}_{\mathrm{T}} \boldsymbol{\Phi}_{\mathrm{T}}^{H} \mathbf{g}_{k_t} }.
\end{equation}
\hrulefill
\end{figure*}

\subsection{Problem Formulation}
We now consider formulating the problem of sum-rate maximization. The objective is jointly design the digital beamformers are the BS, amplification matrix $\mathbf{A}$, the energy splitting matrices $\mathbf{E}_R$ and $\mathbf{E}_T$, and the phase shifting matrices $\mathbf{\Phi}_R$ and $\mathbf{\Phi}_T$. Formally, the optimization problem can be stated as:

\begin{subequations}
\label{prob:SR_BDRIS}
\begin{align}
\max_{\substack{\{\mathbf{w}_k\},\,\boldsymbol{\Phi}_{\mathrm{R}},\,\boldsymbol{\Phi}_{\mathrm{T}},\\
\mathbf{A},\,\mathbf{E}_{\mathrm{R}},\,\mathbf{E}_{\mathrm{T}}}}&
\quad 
\sum_{k_t \in \mathcal{K}_{\mathrm{T}}} \log_2\!\big(1+\gamma_{k_t}\big)
+ \sum_{k_r \in \mathcal{K}_{\mathrm{R}}} \log_2\!\big(1+\gamma_{k_r}\big)
\\[2pt]
\text{s.t.}\quad
& \quad \text{(C1)}-\text{(C6)} 
\\[2pt]
&  \quad \sum_{k \in \mathcal{K}} \mathrm{Tr}\!\big(\mathbf{w}_k \mathbf{w}_k^H\big) \;\le\; P_{\mathrm{BS}},
\label{prob:SR_BDRIS_full_bs} 
\end{align}
\end{subequations}
where $\text{(C1)}-\text{(C6)}$ denote the feasibility 
constraints of the active BD-RIS operation and  \eqref{prob:SR_BDRIS_full_bs} enforces the total transmit power constraint $P_{BS}$ at the BS. 

\section{Proposed Solution} \label{section_3}
\subsection{WMMSE Transformation of the Sum-Rate Maximization Problem}
\label{subsec:wmmse_transformation}

The sum-rate maximization problem in \eqref{prob:SR_BDRIS} is highly non-convex due to the coupled fractional SINR terms $\gamma_{k_r}$ and $\gamma_{k_t}$, which depend nonlinearly on the transmit beamformers, RIS parameters, and amplification factors. To enable efficient optimization, the problem is equivalently reformulated using the WMMSE approach, which converts the non-convex rate expression into a tractable, enabling alternating optimization.

For each reflective-zone user $k_r \in \mathcal{K}_{\mathrm{R}}$ and transmissive-zone user $k_t \in \mathcal{K}_{\mathrm{T}}$, let $u_{k_r}$ and $u_{k_t}$ denote scalar receive equalizers. Their respective mean-square errors (MSEs) can be written as:
\begin{align}
\varepsilon_{k_r} &= \mathbb{E}\!\left[|u_{k_r} y_{k_r} - s_{k_r}|^2\right], \qquad
\varepsilon_{k_t} = \mathbb{E}\!\left[|u_{k_t} y_{k_t} - s_{k_t}|^2\right].
\end{align}
with the following MSE expressions:
\begin{equation}
\label{eq:MSE_reflective}
\begin{aligned}
\varepsilon_{k_r}
&= |u_{k_r}|^{2}\!\Bigg(
\sum_{j\in\mathcal{K}}
\Big|\big(\mathbf{h}_{k_r}^{H} + \mathbf{g}_{k_r}^{H}\boldsymbol{\Phi}_{\mathrm{R}}\mathbf{E}_{\mathrm{R}}\mathbf{A}\mathbf{G}\big)\mathbf{w}_{j}\Big|^{2}
\\&\quad
+\, \sigma_{k_r}^{2}
+\, \mathbf{g}_{k_r}^{H}\boldsymbol{\Phi}_{\mathrm{R}}\mathbf{E}_{\mathrm{R}}\mathbf{A}\mathbf{N}\mathbf{A}\mathbf{E}_{\mathrm{R}}^{H}\boldsymbol{\Phi}_{\mathrm{R}}^{H}\mathbf{g}_{k_r}
\Bigg)
\\&\quad
-\, 2\,\Re\!\left\{
u_{k_r}\,\big(\mathbf{h}_{k_r}^{H} + \mathbf{g}_{k_r}^{H}\boldsymbol{\Phi}_{\mathrm{R}}\mathbf{E}_{\mathrm{R}}\mathbf{A}\mathbf{G}\big)\mathbf{w}_{k_r}
\right\}
+ 1,
\end{aligned}
\end{equation}

\begin{equation}
\label{eq:MSE_transmissive}
\begin{aligned}
\varepsilon_{k_t}
&= |u_{k_t}|^{2}\!\Bigg(
\sum_{j\in\mathcal{K}}
\Big|\big(\mathbf{h}_{k_t}^{H} + \mathbf{g}_{k_t}^{H}\boldsymbol{\Phi}_{\mathrm{T}}\mathbf{E}_{\mathrm{T}}\mathbf{A}\mathbf{G}\big)\mathbf{w}_{j}\Big|^{2}
\\&\quad
+\, \sigma_{k_t}^{2}
+\, \mathbf{g}_{k_t}^{H}\boldsymbol{\Phi}_{\mathrm{T}}\mathbf{E}_{\mathrm{T}}\mathbf{A}\mathbf{N}\mathbf{A}\mathbf{E}_{\mathrm{T}}^{H}\boldsymbol{\Phi}_{\mathrm{T}}^{H}\mathbf{g}_{k_t}
\Bigg)
\\&\quad
-\, 2\,\Re\!\left\{
u_{k_t}\,\big(\mathbf{h}_{k_t}^{H} + \mathbf{g}_{k_t}^{H}\boldsymbol{\Phi}_{\mathrm{T}}\mathbf{E}_{\mathrm{T}}\mathbf{A}\mathbf{G}\big)\mathbf{w}_{k_t}
\right\}
+ 1.
\end{aligned}
\end{equation}
A key property of the WMMSE framework establishes an exact equivalence between the achievable rate $\log_2(1+\gamma_k)$ and a weighted MSE minimization $\forall k \in \mathcal{K}$ \cite{11054266,christensen2008weighted}:
\begin{equation}
\log_2(1+\gamma_k)
= \max_{u_k,\,t_k>0} \log_2(e)\big(\ln t_k - t_k \varepsilon_k(u_k) + 1\big),
\end{equation}
where $t_k>0$ is an auxiliary MSE weight. This identity allows the sum-rate maximization to be reformulated as the following WMMSE-equivalent problem:
\begin{equation}
\label{prob:wmmse_equiv_formulation_concise}
\begin{aligned}
\min_{\substack{\mathbf{w}_{k},
u_{k},t_{k}\\
\boldsymbol{\Phi}_{\mathrm{R}},\,\boldsymbol{\Phi}_{\mathrm{T}},\,\mathbf{A},\,\\\mathbf{E}_{\mathrm{R}},\,\mathbf{E}_{\mathrm{T}}}}
&  
\sum_{k\in\mathcal{K}} (t_{k}\varepsilon_{k} - \ln t_{k})
\\[3pt]
 \text{s.t.}\quad & \text{(C1)}-\text{(C5)}\\
 &\eqref{prob:SR_BDRIS_full_bs}.
\end{aligned}
\end{equation}
where $k = k_r \in \mathcal{K}_R$ or $k = k_t \in \mathcal{K}_T$.
Problem \eqref{prob:wmmse_equiv_formulation_concise} is equivalent in optimality to the original sum-rate problem. This WMMSE reformulation thus provides a structured framework for joint optimization of all system variables based on alternating optimization. The next subsections derives optimal expression for all the variables under the WMMSE paradigm.

\subsection{Optimal MMSE Combiners and MSE Weights}
\label{subsec:wmmse_u_t_joint}

For fixed $\{\mathbf{w}_k\}$, $(\boldsymbol{\Phi}_{\mathrm R}, \boldsymbol{\Phi}_{\mathrm T})$, $\mathbf{A}$, and $(\mathbf{E}_{\mathrm R}, \mathbf{E}_{\mathrm T})$, 
the WMMSE objective in \eqref{prob:wmmse_equiv_formulation_concise} is block-convex in $(u_k, t_k)$. 
Hence, for each user, the scalar MMSE equalizer can obtained by minimizing the corresponding MSE, and the optimal MSE weight follows from maximizing $-\ln t_k + t_k \varepsilon_k$.

Taking the derivative of the MSE expressions in \eqref{eq:MSE_reflective}–\eqref{eq:MSE_transmissive} with respect to the conjugate of $u_k$ yields the closed-form MMSE equalizers for user $k_r \in \mathcal{K}_{\mathrm R}$ and $k_t \in \mathcal{K}_{\mathrm T}$:
\begin{equation}
\label{eq:u_mmse_kr_final_corrected}
\begin{aligned}
u_{k_r}
&= \Big(
\sum\limits_{j\in\mathcal{K}}
\Big|
\big(\mathbf{h}_{k_r}^{H} + \mathbf{g}_{k_r}^{H}\boldsymbol{\Phi}_{\mathrm R}\mathbf{E}_{\mathrm R}\mathbf{A}\mathbf{G}\big)\mathbf{w}_{j}
\Big|^{2}
 \\&+\, \mathbf{g}_{k_r}^{H}\boldsymbol{\Phi}_{\mathrm R}\mathbf{E}_{\mathrm R}\mathbf{A}\mathbf{N}\mathbf{A}^{H}\mathbf{E}_{\mathrm R}^{H}\boldsymbol{\Phi}_{\mathrm R}^{H}\mathbf{g}_{k_r}
+\, \sigma_{k_r}^{2}
\Big)^{-1}
\\& \times \big(\mathbf{h}_{k_r}^{H} + \mathbf{g}_{k_r}^{H}\boldsymbol{\Phi}_{\mathrm R}\mathbf{E}_{\mathrm R}\mathbf{A}\mathbf{G}\big)\mathbf{w}_{k_r}.
\end{aligned}
\end{equation}

\begin{equation}
\label{eq:u_mmse_kt_final_corrected}
\begin{aligned}
u_{k_t}
&= \sum\limits_{j\in\mathcal{K}}
\Big(
\big(\mathbf{h}_{k_t}^{H} + \mathbf{g}_{k_t}^{H}\boldsymbol{\Phi}_{\mathrm T}\mathbf{E}_{\mathrm T}\mathbf{A}\mathbf{G}\big)\mathbf{w}_{j}
\Big|^{2}
\\& +\, \mathbf{g}_{k_t}^{H}\boldsymbol{\Phi}_{\mathrm T}\mathbf{E}_{\mathrm T}\mathbf{A}\mathbf{N}\mathbf{A}^{H}\mathbf{E}_{\mathrm T}^{H}\boldsymbol{\Phi}_{\mathrm T}^{H}\mathbf{g}_{k_t}
+\, \sigma_{k_t}^{2}\Big)^{-1}
\\&  \times \big(\mathbf{h}_{k_t}^{H} + \mathbf{g}_{k_t}^{H}\boldsymbol{\Phi}_{\mathrm T}\mathbf{E}_{\mathrm T}\mathbf{A}\mathbf{G}\big)\mathbf{w}_{k_t}
\end{aligned}
\end{equation}
Let $\varepsilon_{k_r}^{\mathrm{MMSE}}$ and $\varepsilon_{k_t}^{\mathrm{MMSE}}$ denote the corresponding MMSE obtained by substituting the optimal combiners in the MSE expression.  Then, the optimal MSE weights corresponding to these minimum errors are given by \cite{11054266,christensen2008weighted}:
\begin{equation}
\label{eq:t_mmse_kr_final_corrected}
t_{k_r} = \frac{1}{\varepsilon_{k_r}^{\mathrm{MMSE}}},
\qquad
t_{k_t} = \frac{1}{\varepsilon_{k_t}^{\mathrm{MMSE}}}.
\end{equation}
The expressions 
\eqref{eq:u_mmse_kr_final_corrected}–\eqref{eq:t_mmse_kr_final_corrected}
provide closed-form updates for the scalar MMSE combiners and MSE weights of the reflective and transmissive users, respectively.

\subsection{WMMSE-Based Digital Beamforming}
\label{subsec:wmmse_beamforming_exact}
The digital beamforming stage at the BS aims to jointly design the transmit beamformers for all users located in both the reflective and transmissive zones of the BD-RIS-assisted system. 
Given the MMSE receivers $\{u_{k_r},u_{k_t}\}$ and the MSE weights $\{t_{k_r},t_{k_t}\}$, assuming the other variables fixed, the transmit-side subproblem can be expressed as:
\begin{equation}
\label{eq:wmmse_w_subprob_exact}
\begin{aligned}
\min_{\{\mathbf{w}_{k_r}\},\{\mathbf{w}_{k_t}\}} \quad &
\sum_{k_r\in\mathcal{K}_{\mathrm{R}}} t_{k_r}\, \varepsilon_{k_r}(\mathbf{w}_{k_r})
+ \sum_{k_t\in\mathcal{K}_{\mathrm{T}}} t_{k_t}\, \varepsilon_{k_t}(\mathbf{w}_{k_t}) \\[4pt]
\text{s.t.}\quad &
\sum_{k_r\in\mathcal{K}_{\mathrm{R}}} \hspace{-1mm} \mathrm{Tr}\!\big(\mathbf{w}_{k_r} \mathbf{w}_{k_r}^H\big) 
+ \sum_{k_t\in\mathcal{K}_{\mathrm{T}}} \hspace{-1mm} \mathrm{Tr}\!\big(\mathbf{w}_{k_t} \mathbf{w}_{k_t}^H\big) \hspace{-1mm}  \le \hspace{-1mm} P_{\mathrm{BS}},
\end{aligned}
\end{equation}
where $P_{\mathrm{BS}}$ denotes the maximum total transmit power at the BS. 
The objective in \eqref{eq:wmmse_w_subprob_exact} is quadratic and convex with respect to $\{\mathbf{w}_{k_r}\}$ and $\{\mathbf{w}_{k_t}\}$ for fixed $\{u_k,t_k\}$, ensuring a unique global minimizer.

Using the MSE expressions computed with the optimal MMSE equalizers and the MSE weights in \eqref{eq:MSE_reflective}–\eqref{eq:MSE_transmissive}, the objective can be expanded as
\begin{equation}
\label{eq:wmmse_expanded_objective}
\begin{aligned}
\mathcal{J}(\{\mathbf{w}_k\})
&= \sum_{k_r\in\mathcal{K}_{\mathrm R}} 
t_{k_r}\!\Big(
|u_{k_r}|^{2}\!\sum_{j\in\mathcal{K}}
\Big|\big(\mathbf{h}_{k_r}^{H}\!+\!\mathbf{g}_{k_r}^{H}\boldsymbol{\Phi}_{\mathrm R}\mathbf{E}_{\mathrm R}\mathbf{A}\mathbf{G}\big)\mathbf{w}_{j}\Big|^{2}
\\&\quad
-2\Re\!\left\{u_{k_r}\!\big(\mathbf{h}_{k_r}^{H}\!+\!\mathbf{g}_{k_r}^{H}\boldsymbol{\Phi}_{\mathrm R}\mathbf{E}_{\mathrm R}\mathbf{A}\mathbf{G}\big)\mathbf{w}_{k_r}\right\}
\\&+ |u_{k_r}|^{2}\!\Big(\mathbf{g}_{k_r}^{H}\boldsymbol{\Phi}_{\mathrm R}\mathbf{E}_{\mathrm R}\mathbf{A}\mathbf{N}\mathbf{A}^{H}\mathbf{E}_{\mathrm R}^{H}\boldsymbol{\Phi}_{\mathrm R}^{H}\mathbf{g}_{k_r}\!+\!\sigma_{k_r}^{2}\Big)
\!+\!1
\Big)
\\&\quad
+\sum_{k_t\in\mathcal{K}_{\mathrm T}} 
t_{k_t}\!\Big(
|u_{k_t}|^{2}\!\sum_{j\in\mathcal{K}}
\Big|\big(\mathbf{h}_{k_t}^{H}\!+\!\mathbf{g}_{k_t}^{H}\boldsymbol{\Phi}_{\mathrm T}\mathbf{E}_{\mathrm T}\mathbf{A}\mathbf{G}\big)\mathbf{w}_{j}\Big|^{2}
\\&\quad
-2\Re\!\left\{u_{k_t}\!\big(\mathbf{h}_{k_t}^{H}\!+\!\mathbf{g}_{k_t}^{H}\boldsymbol{\Phi}_{\mathrm T}\mathbf{E}_{\mathrm T}\mathbf{A}\mathbf{G}\big)\mathbf{w}_{k_t}\right\}
\\& + |u_{k_t}|^{2}\!\Big(\mathbf{g}_{k_t}^{H}\boldsymbol{\Phi}_{\mathrm T}\mathbf{E}_{\mathrm T}\mathbf{A}\mathbf{N}\mathbf{A}^{H}\mathbf{E}_{\mathrm T}^{H}\boldsymbol{\Phi}_{\mathrm T}^{H}\mathbf{g}_{k_t}\!+\!\sigma_{k_t}^{2}\Big)
\!+\!1
\Big).
\end{aligned}
\end{equation}
Differentiating the Lagrangian associated with \eqref{eq:wmmse_w_subprob_exact} with respect to $\mathbf{w}_{k_r}$ and $\mathbf{w}_{k_t}$ and setting the gradients to zero yields the KKT stationarity conditions. 
Let $\lambda \ge 0$ denote the dual variable associated with the BS power constraint. Then, for reflective and transmissive users, we have the following KKT conditions:
\begin{align}
\big(\mathbf{Q} + \lambda \mathbf{I}_M\big)\mathbf{w}_{k_r}
&= t_{k_r} u_{k_r}
\Big(\mathbf{h}_{k_r} + \mathbf{G}^{H}\mathbf{A}^{H}\mathbf{E}_{\mathrm R}^{H}\boldsymbol{\Phi}_{\mathrm R}^{H}\mathbf{g}_{k_r}\Big),
\label{eq:reflective_stationarity_corrected}
\\[4pt]
\big(\mathbf{Q} + \lambda \mathbf{I}_M\big)\mathbf{w}_{k_t}
&= t_{k_t} u_{k_t}
\Big(\mathbf{h}_{k_t} + \mathbf{G}^{H}\mathbf{A}^{H}\mathbf{E}_{\mathrm T}^{H}\boldsymbol{\Phi}_{\mathrm T}^{H}\mathbf{g}_{k_t}\Big),
\label{eq:transmissive_stationarity_corrected}
\end{align}
where $\mathbf{Q}$ denotes the aggregate curvature matrix of the WMMSE objective, capturing all multiuser interference terms. It is given as $\label{eq:Q_decomposition_corrected}
\mathbf{Q} = \mathbf{Q}_{\mathrm R} + \mathbf{Q}_{\mathrm T},$
where
\begin{equation}
\label{eq:Q_R_corrected}
\begin{aligned}
    \mathbf{Q}_{\mathrm R}
=  \sum_{k_r\in\mathcal{K}_{\mathrm R}} t_{k_r} |u_{k_r}|^{2}
\Big(&\mathbf{h}_{k_r} + \mathbf{G}^{H}\mathbf{A}^{H}\mathbf{E}_{\mathrm R}^{H}\boldsymbol{\Phi}_{\mathrm R}^{H}\mathbf{g}_{k_r}\Big) \\& \times 
\Big(\mathbf{h}_{k_r} + \mathbf{G}^{H}\mathbf{A}^{H}\mathbf{E}_{\mathrm R}^{H}\boldsymbol{\Phi}_{\mathrm R}^{H}\mathbf{g}_{k_r}\Big)^{\!H},
\end{aligned}
\end{equation}
\begin{equation}
\label{eq:Q_T_corrected}
\begin{aligned}
\mathbf{Q}_{\mathrm T}
= \sum_{k_t\in\mathcal{K}_{\mathrm T}} t_{k_t} |u_{k_t}|^{2}
\Big(& \mathbf{h}_{k_t} + \mathbf{G}^{H}\mathbf{A}^{H}\mathbf{E}_{\mathrm T}^{H}\boldsymbol{\Phi}_{\mathrm T}^{H}\mathbf{g}_{k_t}\Big) \\& \times
\Big(\mathbf{h}_{k_t} + \mathbf{G}^{H}\mathbf{A}^{H}\mathbf{E}_{\mathrm T}^{H}\boldsymbol{\Phi}_{\mathrm T}^{H}\mathbf{g}_{k_t}\Big)^{\!H}.
\end{aligned}
\end{equation}

Each term $t_k |u_k|^2 (\cdot)(\cdot)^H$ is Hermitian and positive semidefinite, ensuring $\mathbf{Q} \succeq \mathbf{0}$. Thus, the subproblem \eqref{eq:wmmse_w_subprob_exact} is convex in $\{\mathbf{w}_k\}, \forall k$. The closed-form optimal beamforming vectors are then obtained by directly solving 
\eqref{eq:reflective_stationarity_corrected}–\eqref{eq:transmissive_stationarity_corrected}, leading to the following closed form solution:
\begin{align}
\label{eq:wkr_optimal_corrected}
\mathbf{w}_{k_r}
&= \big(\mathbf{Q} + \lambda \mathbf{I}_M\big)^{-1}
t_{k_r} u_{k_r}^{*}
\Big(\mathbf{h}_{k_r} + \mathbf{G}^{H}\mathbf{A}^{H}\mathbf{E}_{\mathrm R}^{H}\boldsymbol{\Phi}_{\mathrm R}^{H}\mathbf{g}_{k_r}\Big), \\[4pt]
\label{eq:wkt_optimal_corrected}
\mathbf{w}_{k_t}
&= \big(\mathbf{Q} + \lambda \mathbf{I}_M\big)^{-1}
t_{k_t} u_{k_t}^{*}
\Big(\mathbf{h}_{k_t} + \mathbf{G}^{H}\mathbf{A}^{H}\mathbf{E}_{\mathrm T}^{H}\boldsymbol{\Phi}_{\mathrm T}^{H}\mathbf{g}_{k_t}\Big).
\end{align}
Finally, the dual variable $\lambda$ regulates the total BS transmit power and is chosen such that
\begin{equation}
\sum_{k_r\in\mathcal{K}_{\mathrm{R}}} \hspace{-1mm} \mathrm{Tr}\!\big(\mathbf{w}_{k_r} \mathbf{w}_{k_r}^H\big) 
+ \sum_{k_t\in\mathcal{K}_{\mathrm{T}}} \hspace{-1mm} \mathrm{Tr}\!\big(\mathbf{w}_{k_t} \mathbf{w}_{k_t}^H\big) \hspace{-1mm} =   P_{\mathrm{BS}},
\end{equation}
which can be efficiently solved via a one-dimensional bisection search due to the monotonic relationship between total transmit power and $\lambda$.

\subsection{WMMSE-Based Amplification Matrix Optimization}
\label{subsec:wmmse_amplification}

We consider optimizing the diagonal amplification matrix
\(\mathbf{A}=\operatorname{diag}(\boldsymbol{\beta})\), where
\(\boldsymbol{\beta}=[\beta_1,\ldots,\beta_N]^T \in \mathbb{R}_{\ge 0}^N\) and
\(\beta_i=\sqrt{a_i}\) denotes the amplitude gain of element \(i\).
For fixed digital precoders \(\{\mathbf{w}_k\}\), MMSE combiners \(\{u_k\}\),
weights \(\{t_k\}\), coupling \(\boldsymbol{\Phi}_{\mathrm R},\boldsymbol{\Phi}_{\mathrm T}\),
and splitters \(\mathbf{E}_{\mathrm R},\mathbf{E}_{\mathrm T}\), the WMMSE subproblem in \(\mathbf{A}\) reads
\begin{subequations}\label{prob:wmmse_A_correct}
\begin{equation}
\min_{\boldsymbol{\beta}\ge 1}\quad
 \sum_{k_r\in\mathcal{K}_{\mathrm R}} t_{k_r}\,\varepsilon_{k_r}(\boldsymbol{\beta})
 + \sum_{k_t\in\mathcal{K}_{\mathrm T}} t_{k_t}\,\varepsilon_{k_t}(\boldsymbol{\beta}),
\label{eq:wmmse_A_total_power_correct}
\end{equation}
\begin{equation}
\begin{aligned}
    \text{s.t.}\quad
& 
\operatorname{diag}\!\Big(
\boldsymbol{\Phi}_{\mathrm R}\,\mathbf E_{\mathrm R}\,\mathbf A\,\boldsymbol{\Sigma}_v\,\mathbf A\,\mathbf E_{\mathrm R}\,\boldsymbol{\Phi}_{\mathrm R}^{H} \\&+
\boldsymbol{\Phi}_{\mathrm T}\,\mathbf E_{\mathrm T}\,\mathbf A\,\boldsymbol{\Sigma}_v\,\mathbf A\,\mathbf E_{\mathrm T}\,\boldsymbol{\Phi}_{\mathrm T}^{H}
\Big)\ \le\ \mathbf P_{\max}, \label{eq:per_element_cap_restated}
\end{aligned}
\end{equation}
\begin{equation}
\begin{aligned}
\operatorname{Tr}\!\Big(&
\boldsymbol{\Phi}_{\mathrm R}\,\mathbf E_{\mathrm R}\,\mathbf A\,\boldsymbol{\Sigma}_v\,\mathbf A\,\mathbf E_{\mathrm R}\,\boldsymbol{\Phi}_{\mathrm R}^{H} \\&+\boldsymbol{\Phi}_{\mathrm T}\,\mathbf E_{\mathrm T}\,\mathbf A\,\boldsymbol{\Sigma}_v\,\mathbf A\,\mathbf E_{\mathrm T}\,\boldsymbol{\Phi}_{\mathrm T}^{H}\Big)\le\ P_{\max}, \label{eq:total_cap}
    \end{aligned}
\end{equation}
\end{subequations}
where $\boldsymbol{\Sigma}_v=\mathbf G(\sum_{k\in\mathcal K}\mathbf w_k\mathbf w_k^H)\mathbf G^H + \mathbf{N}$ is the BS--RIS forwarded signal covariance, $\mathbf P_{\max}\in\mathbb R_+^N$ caps the per-element emitted powers at the two RIS sides jointly, and $P_{\max}$ is the total emission budget. This is a quadratic optimization problem, and can be solved with any convex solver, e.g. CVX.

 \subsection{WMMSE-Driven Optimal Power Splitting}
\label{subsec:wmmse_ER_ET}

Recall that $\mathbf{E}_{\mathrm{R}} \triangleq \operatorname{diag}(\varsigma_1, \dots, \varsigma_N)$ denote the reflective-zone power-splitting matrix, where $\varsigma_i = \sqrt{\alpha_i} \in [0,1]$ represents the amplitude reflection coefficient of the $i$th element. The corresponding transmissive-zone matrix is defined as
$\mathbf{E}_{\mathrm{T}} \triangleq \operatorname{diag}\big(\sqrt{1 - \varsigma_1^2}, \dots, \sqrt{1 - \varsigma_N^2}\big)$. For fixed $\{u_k, t_k\}$, $\{\mathbf{w}_k\}$, $\mathbf{A}$, and $(\boldsymbol{\Phi}_{R},\boldsymbol{\Phi}_{T})$, the WMMSE subproblem for the joint optimization of $\mathbf{E}_{\mathrm{R}}$ and $\mathbf{E}_{\mathrm{T}}$ can be formulated as
\begin{subequations}
\label{prob:ERET_wmmse}
\begin{align}
\min_{\mathbf{E}_{\mathrm{R}}, \mathbf{E}_{\mathrm{T}}}\quad & \sum_{k\in\mathcal{K}} t_k\, \varepsilon_k(\mathbf{E}_{\mathrm{R}}, \mathbf{E}_{\mathrm{T}}) \\[2pt]
\text{s.t.}\quad & 0 \le \varsigma_i \le 1,\quad i=1,\ldots,N,\\[2pt]
& \mathbf{E}_{\mathrm{R}}^{H}\mathbf{E}_{\mathrm{R}} + \mathbf{E}_{\mathrm{T}}^{H}\mathbf{E}_{\mathrm{T}} = \mathbf{I}_N. \label{eq:ERET_energy_constraint}
\end{align}
\end{subequations}
To solve this constrained optimization problem efficiently while guaranteeing monotonic convergence, we propose a cyclic coordinate descent algorithm with global acceptance criteria. 
The weighted MSE objective function admits a quadratic decomposition in terms of the diagonal elements of $\mathbf{E}_{\mathrm{R}}$ and $\mathbf{E}_{\mathrm{T}}$. Let $\mathbf{e}_{\mathrm{R}} = \operatorname{diag}(\mathbf{E}_{\mathrm{R}}) = \boldsymbol{\varsigma}$ and $\mathbf{e}_{\mathrm{T}} = \operatorname{diag}(\mathbf{E}_{\mathrm{T}}) = \sqrt{\mathbf{1} - \boldsymbol{\varsigma} \odot \boldsymbol{\varsigma}}$. The objective function can be expressed as:

\begin{equation}
\label{eq:ERET_quadratic_form}
\begin{aligned}
\sum_{k\in\mathcal{K}} t_k\, \varepsilon_k(\boldsymbol{\varsigma})
= &\ \boldsymbol{\varsigma}^{H}\mathbf{S}_{\mathrm{R}}\boldsymbol{\varsigma}
- 2\,\Re\{\mathbf{p}_{\mathrm{R}}^{H}\boldsymbol{\varsigma}\}
+ (\sqrt{\mathbf{1}-\boldsymbol{\varsigma}\odot\boldsymbol{\varsigma}})^{H}\mathbf{S}_{\mathrm{T}}\\& \times (\sqrt{\mathbf{1}-\boldsymbol{\varsigma}\odot\boldsymbol{\varsigma}})  - 2\,\Re\{\mathbf{p}_{\mathrm{T}}^{H}(\sqrt{\mathbf{1}-\boldsymbol{\varsigma}\odot\boldsymbol{\varsigma}})\}
+ c,
\end{aligned}
\end{equation}
where $c$ represents constant terms independent of $\boldsymbol{\varsigma}$, and the coefficient matrices are defined as:

\begin{equation}    \label{eq:QR_definition}
\begin{aligned}
\mathbf{S}_{\mathrm{R}} =  \sum_{k_r\in\mathcal{K}_{\mathrm{R}}} t_{k_r} |u_{k_r}|^2 \Big[ &\left(\boldsymbol{\Phi}_{\mathrm{R}}^{H}\mathbf{g}_{k_r}\mathbf{g}_{k_r}^{H}\boldsymbol{\Phi}_{\mathrm{R}}\right)  \odot \Big(\mathbf{A}\mathbf{G}\mathbf{W}\mathbf{W}^{H} \\& \times \mathbf{G}^{H}\mathbf{A}^{H} + \mathbf{A}\mathbf{N}\mathbf{A}^{H}\Big) \Big],
\end{aligned}
\end{equation}

\begin{equation}
\label{eq:QT_definition}
\begin{aligned}
    \mathbf{S}_{\mathrm{T}} = \sum_{k_t\in\mathcal{K}_{\mathrm{T}}} t_{k_t} |u_{k_t}|^2 \Big[ & \left(\boldsymbol{\Phi}_{\mathrm{T}}^{H}\mathbf{g}_{k_t}\mathbf{g}_{k_t}^{H}\boldsymbol{\Phi}_{\mathrm{T}}\right)  \odot \Big(\mathbf{A}\mathbf{G}\mathbf{W}\mathbf{W}^{H} \\& \times \mathbf{G}^{H}\mathbf{A}^{H} + \mathbf{A}\mathbf{N}\mathbf{A}^{H}\Big) \Big],
\end{aligned}
\end{equation}
with the linear coefficient vectors given by:
\begin{equation}
\label{eq:pR_definition}
\mathbf{p}_{\mathrm{R}} = \sum_{k_r\in\mathcal{K}_{\mathrm{R}}} t_{k_r} u_{k_r} \left[ \left(\boldsymbol{\Phi}_{\mathrm{R}}^{H}\mathbf{g}_{k_r}\right) \odot \left(\mathbf{A}\mathbf{G}\mathbf{w}_{k_r}\right) \right],
\end{equation}

\begin{equation}
\label{eq:pT_definition}
\mathbf{p}_{\mathrm{T}} = \sum_{k_t\in\mathcal{K}_{\mathrm{T}}} t_{k_t} u_{k_t} \left[ \left(\boldsymbol{\Phi}_{\mathrm{T}}^{H}\mathbf{g}_{k_t}\right) \odot \left(\mathbf{A}\mathbf{G}\mathbf{w}_{k_t}\right) \right].
\end{equation}
The proposed algorithm employs cyclic coordinate descent, where each element $\varsigma_m$ is optimized sequentially while keeping all other elements fixed. For each element $m$, the dependence of the objective function on $\varsigma_m$ can be isolated as:
\begin{equation}
\label{eq:element_wise_objective}
f_m(\varsigma_m) = s_{r,m}\,\varsigma_m^2 + r_m\,\varsigma_m + s_{t,m}\,(1-\varsigma_m^2) + t_m\,\sqrt{1-\varsigma_m^2},
\end{equation}

where the coefficients are computed from the full matrices as:

\begin{subequations}
\label{eq:local_coefficients}
\begin{align}
s_{r,m} &= [\mathbf{S}_{\mathrm{R}}]_{mm}, \\
s_{t,m} &= [\mathbf{S}_{\mathrm{T}}]_{mm}, \\
r_m &= 2\,\Re\left\{\sum_{n\neq m} [\mathbf{S}_{\mathrm{R}}]_{mn}\,\varsigma_n - [\mathbf{p}_{\mathrm{R}}]_m\right\}, \\
t_m &= 2\,\Re\left\{\sum_{n\neq m} [\mathbf{S}_{\mathrm{T}}]_{mn}\,\sqrt{1-\varsigma_n^2} - [\mathbf{p}_{\mathrm{T}}]_m\right\}.
\end{align}
\end{subequations}
To ensure monotonic convergence while exploring continuous solutions away from trivial boundaries, we generate a comprehensive set of candidate values for each element update. The candidate set $\mathcal{S}_m$ for element $m$ is constructed as:

\begin{equation}
\label{eq:candidate_set}
\mathcal{S}_m = \{0, 1, \varsigma_m^{(k)}\} \cup \mathcal{S}_m^{\text{int}},
\end{equation}
where $\varsigma_m^{(k)}$ denotes the current value at iteration $k$, and $\mathcal{S}_m^{\text{int}}$ represents the interior candidate set containing strategically chosen points. The construction of $\mathcal{S}_m^{\text{int}}$ is detailed as follows:
\begin{subequations}
\label{eq:interior_candidates}
\begin{align}
\mathcal{S}_m^{\text{quad}} &= \left\{ -\frac{r_m}{2(s_{r,m} - s_{t,m})} \right\} \quad \text{if } s_{r,m} \neq s_{t,m} \text{ and} \in (0.1, 0.9), \\
\mathcal{S}_m^{\text{bal}} &= \left\{ 1 - \frac{s_{t,m}}{s_{r,m} + s_{t,m}} \right\} \quad \text{if } s_{r,m}, s_{t,m} > 0, \\
\mathcal{S}_m^{\text{local}} &= \left\{ \varsigma_m^{(k)} + \delta : \delta \in \{-0.2, -0.1, 0.1, 0.2\} \right\} \cap (0.1, 0.9).
\end{align}
\end{subequations}
Each subset in $\mathcal{S}_m^{\text{int}}$ serves a distinct purpose in the optimization strategy:

\begin{itemize}
\item \emph{Quadratic Stationary Point ($\mathcal{S}_m^{\text{quad}}$)}: This candidate represents the stationary point of the quadratic approximation of the objective function, obtained by neglecting the square root term in \eqref{eq:element_wise_objective}. The expression $-\frac{r_m}{2(s_{r,m} - s_{t,m})}$ is derived by setting the derivative of the quadratic part to zero. This candidate captures the optimal solution when the square root term has minimal influence, providing a mathematically grounded exploration point.

\item \emph{Balance Point ($\mathcal{S}_m^{\text{bal}}$)}: This candidate implements a ratio-based approach that balances the relative strengths of reflective and transmissive coefficients. The expression $1 - \frac{s_{t,m}}{s_{r,m} + s_{t,m}}$ ensures that when $s_{r,m} \gg s_{t,m}$, the candidate approaches 1 (prioritizing reflection), and when $s_{t,m} \gg s_{r,m}$, it approaches 0 (prioritizing transmission). This heuristic promotes solutions that adapt to the relative importance of each domain.

\item \emph{Local Exploration ($\mathcal{S}_m^{\text{local}}$)}: This set performs local search around the current value $\varsigma_m^{(k)}$ using perturbations $\delta \in \{-0.2, -0.1, 0.1, 0.2\}$. The chosen step sizes provide both coarse ($\pm 0.2$) and fine ($\pm 0.1$) exploration around the current iterate, enabling the algorithm to escape shallow local minima while maintaining search refinement.
\end{itemize}
The conditional inclusion criteria for $\mathcal{S}_m^{\text{quad}}$ and $\mathcal{S}_m^{\text{bal}}$ ensure numerical stability and feasibility. The restriction to $(0.1, 0.9)$ in $\mathcal{S}_m^{\text{quad}}$ prevents boundary dominance, while the positivity condition in $\mathcal{S}_m^{\text{bal}}$ ensures meaningful ratio computation. The critical point that guarantees monotonic convergence is the global acceptance criterion. For each candidate $\varsigma_m^{\text{cand}} \in \mathcal{S}_m$, we compute the hypothetical global objective function:
\begin{equation}
\label{eq:global_test}
f^{\text{test}} = \sum_{k\in\mathcal{K}} t_k\, \varepsilon_k(\boldsymbol{\varsigma}^{\text{test}}),
\end{equation}
where $\boldsymbol{\varsigma}^{\text{test}}$ is the vector with all elements fixed at their current values except $\varsigma_m$ set to $\varsigma_m^{\text{cand}}$. Accounting for the global objective function is necessary due to mutual coupling. The update is accepted only if:
\begin{equation}
\label{eq:acceptance_condition}
f^{\text{test}} < f(\boldsymbol{\varsigma}^{(k)}),
\end{equation}
where $f(\boldsymbol{\varsigma}^{(k)})$ is the current global objective value. The overall optimal energy splitting algorithm is formally stated in Algorithm~\ref{alg:energy_splitting}, which generates the sequence of objective values $\{f(\boldsymbol{\varsigma}^{(k)})\}$ which is non-increasing and bounded below by zero. The comprehensive candidate set construction ensures efficient exploration of the solution space while the global acceptance criterion maintains strict monotonicity throughout the optimization process, assuring convergence.

 \begin{algorithm}[t]
\caption{Optimal Energy Splitting for Active STAR BD-RIS}
\label{alg:energy_splitting}
\begin{algorithmic}[1]
\State Initialize $\boldsymbol{\varsigma}^{(0)} \in [0.3, 0.7]^N$
\State Precompute $\mathbf{S}_{\mathrm{R}}, \mathbf{S}_{\mathrm{T}}, \mathbf{p}_{\mathrm{R}}, \mathbf{p}_{\mathrm{T}}$
\For{$k = 0, 1, \ldots$ until convergence}
    \State $f_{\text{curr}} \gets \sum_k t_k \varepsilon_k(\boldsymbol{\varsigma}^{(k)})$
    \For{$m = 1$ to $N$}
        \State Compute $s_{r,m}, s_{t,m}, r_m, t_m$
        \State $\mathcal{S}_m \gets \{0, 1, \varsigma_m^{(k)}\} \cup \mathcal{S}_m^{\text{int}}$
        \State $\varsigma_m^{\text{best}} \gets \varsigma_m^{(k)}$, $f_{\text{best}} \gets f_{\text{curr}}$
        \For{$\varsigma_m^{\text{cand}} \in \mathcal{S}_m$}
            \State $\boldsymbol{\varsigma}^{\text{test}} \gets \boldsymbol{\varsigma}^{(k)}$ with $\varsigma_m = \varsigma_m^{\text{cand}}$
            \State $f^{\text{test}} \gets \sum_k t_k \varepsilon_k(\boldsymbol{\varsigma}^{\text{test}})$
            \If{$f^{\text{test}} < f_{\text{best}}$}
                \State $\varsigma_m^{\text{best}} \gets \varsigma_m^{\text{cand}}$, $f_{\text{best}} \gets f^{\text{test}}$
            \EndIf
        \EndFor
        \State $\varsigma_m^{(k+1)} \gets \varsigma_m^{\text{best}}$
    \EndFor
    \If{converged} \textbf{break} \EndIf
\EndFor
\State \textbf{return} $\mathbf{E}_{\mathrm{R}}^*, \mathbf{E}_{\mathrm{T}}^*$, with $\boldsymbol{\varsigma}^*$.
\end{algorithmic}
\end{algorithm}
 
\subsection{WMMSE-Driven Optimization of BD-RIS}
\label{subsec:wmmse_bd_ris_formal}

This subsection derives the optimal BD-RIS responses, represented by the reflective and transmissive coupling matrices $\boldsymbol{\Phi}_{\mathrm{R}}$ and $\boldsymbol{\Phi}_{\mathrm{T}}$, within the WMMSE framework. These matrices govern the electromagnetic behavior of the BD-RIS, jointly shaping the reflected and transmitted fields to minimize the system-wide weighted sum of MSEs. For given MMSE equalizers $\{u_k\}$, user weights $\{t_k\}$, digital precoders $\{\mathbf{w}_k\}$, amplification matrix $\mathbf{A}$, and power-splitting matrices $(\mathbf{E}_{\mathrm{R}}, \mathbf{E}_{\mathrm{T}})$, the optimization sub-problem is formulated as:
\begin{subequations}
\label{prob:BDRIS}
\begin{equation}
\begin{aligned}
\min_{\boldsymbol{\Phi}_{\mathrm{R}},\,\boldsymbol{\Phi}_{\mathrm{T}}}\;
&\sum_{k_r\in\mathcal{K}_{\mathrm{R}}}
t_{k_r}\,\varepsilon_{k_r}(\boldsymbol{\Phi}_{\mathrm{R}},\boldsymbol{\Phi}_{\mathrm{T}})
+\sum_{k_t\in\mathcal{K}_{\mathrm{T}}}
t_{k_t}\,\varepsilon_{k_t}(\boldsymbol{\Phi}_{\mathrm{R}},\boldsymbol{\Phi}_{\mathrm{T}}),
\end{aligned}
\end{equation}
\begin{equation} 
    \text{s.t.} \quad 
    \boldsymbol{\Phi}_{\mathrm T}^{H} \boldsymbol{\Phi}_{\mathrm R} = \bmI, \& \quad
\boldsymbol{\Phi}_{\mathrm T}^{H} \boldsymbol{\Phi}_{\mathrm R} = \bmI
\end{equation}
\end{subequations}
The optimization problem in \eqref{prob:BDRIS} can be decomposed into two independent subproblems, each subject to separate constraints that ensure the conservation of total amplified and split energy. Notably, since the coupling between $\mathbf{\Phi}_{T}$ and $\mathbf{\Phi}_{R}$ has been removed, they can be optimized in a distributed and parallel manner while keeping the other variables fixed.\footnote{It is worth noting that, as the overall algorithmic complexity is primarily determined by the updates of $\boldsymbol{\Phi}_{R}$ and $\boldsymbol{\Phi}_{T}$, therefore the distributed implementation offers significant advantages for real-time applications as it enables the optimization of the transmissive and reflective responses in parallel.}

\subsubsection{Optimization of $\boldsymbol{\Phi}_R$} 
We optimize the reflective response by considering the following sub-problem
\begin{subequations}
\begin{equation}
\min_{\boldsymbol{\Phi}_{\mathrm R}}
\ \sum_{k_r\in\mathcal{K}_{\mathrm R}}
t_{k_r}\,\varepsilon_{k_r}(\boldsymbol{\Phi}_{\mathrm R},\boldsymbol{\Phi}_{\mathrm T})
\end{equation}
\vspace{-0.5em}
\begin{equation}
\text{s.t.}\quad
\boldsymbol{\Phi}_{\mathrm R}^{H}\boldsymbol{\Phi}_{\mathrm R}=\mathbf I ,
\end{equation}
\end{subequations}
which constrains $\boldsymbol{\Phi}_{\mathrm R}$ to the complex Stiefel manifold \cite{nerini2023closed,li2023beyond}. Let $\mathbf W\!=\![\mathbf w_1,\ldots,\mathbf w_{K}]$ and define the fixed (within this block) matrices
\begin{subequations}
    \begin{equation}
        \mathbf S_{\mathrm R}\triangleq \mathbf E_{\mathrm R}\mathbf A\mathbf G\mathbf W, \quad
        \mathbf C_{\mathrm R}\triangleq \mathbf E_{\mathrm R}\mathbf A\mathbf N\mathbf A\mathbf E_{\mathrm R}^{H},
    \end{equation}
    \begin{equation}
        \mathbf Q_{\mathrm R}\triangleq \mathbf S_{\mathrm R}\mathbf S_{\mathrm R}^{H}+\mathbf C_{\mathrm R}.
    \end{equation}
\end{subequations}
For each $k_r\!\in\!\mathcal K_{\mathrm R}$, let $d_{k_r,j}\!\triangleq\!\mathbf h_{k_r}^{H}\mathbf w_j$, let $\mathbf s_{\mathrm R,j}$ denote the $j$-th column of $\mathbf S_{\mathrm R}$, and define $v_{k_r,\mathrm R}$ as
\begin{equation}
\mathbf v_{k_r,\mathrm R}\triangleq |u_{k_r}|^{2}\sum_{j\in\mathcal K}\mathbf s_{\mathrm R,j}\,d_{k_r,j}^{*}\;-\;u_{k_r}\,\mathbf s_{\mathrm R,k_r}.
\end{equation}
Discarding terms independent of $\boldsymbol{\Phi}_{\mathrm R}$, the objective function can be rewritten as
\begin{equation}
    \begin{aligned}
      f_{\mathrm R}(\boldsymbol{\Phi}_{\mathrm R})
=&\sum_{k_r\in\mathcal K_{\mathrm R}} t_{k_r}\!\Big(
|u_{k_r}|^{2}\,\mathrm{tr}\!\big(\boldsymbol{\Phi}_{\mathrm R}^{H}\mathbf g_{k_r}\mathbf g_{k_r}^{H}\boldsymbol{\Phi}_{\mathrm R}\mathbf Q_{\mathrm R}\big)
\\&+ 2\,\Re\!\{\mathrm{tr}(\mathbf v_{k_r,\mathrm R}\mathbf g_{k_r}^{H}\boldsymbol{\Phi}_{\mathrm R})\}
\Big).  
    \end{aligned}
\end{equation}
The Euclidean gradient, using $\frac{\partial}{\partial \mathbf X}\mathrm{tr}(\mathbf X^{H}\mathbf A\mathbf X\mathbf B)=\mathbf A\mathbf X\mathbf B$ and $\nabla_{\mathbf X}\,2\Re\{\mathrm{tr}(\mathbf C^{H}\mathbf X)\}=2\mathbf C$, is given as
\begin{equation}
\nabla f_{\mathrm R}(\boldsymbol{\Phi}_{\mathrm R})
=\sum_{k_r} t_{k_r}\!\left(
|u_{k_r}|^{2}\,\mathbf g_{k_r}\mathbf g_{k_r}^{H}\,\boldsymbol{\Phi}_{\mathrm R}\,\mathbf Q_{\mathrm R}
+ 2\,\mathbf g_{k_r}\mathbf v_{k_r,\mathrm R}^{H}
\right).
\end{equation}
To simplify notations, we aggregate constants terms as
\begin{subequations}
    \begin{equation}
        \mathbf M_{\mathrm R}\triangleq\sum_{k_r} t_{k_r}|u_{k_r}|^{2}\,\mathbf g_{k_r}\mathbf g_{k_r}^{H},\qquad
\mathbf B_{\mathrm R}\triangleq\sum_{k_r} 2t_{k_r}\,\mathbf g_{k_r}\mathbf v_{k_r,\mathrm R}^{H},
    \end{equation}
    and the gradient can be expressed in a compact form as
    \begin{equation}
        \nabla f_{\mathrm R}(\boldsymbol{\Phi}_{\mathrm R})=\mathbf M_{\mathrm R}\boldsymbol{\Phi}_{\mathrm R}\mathbf Q_{\mathrm R}+\mathbf B_{\mathrm R}
    \end{equation}
\end{subequations}
Projecting onto the tangent space $\mathcal T_{\boldsymbol{\Phi}_{\mathrm R}}\mathrm{St}=\{\boldsymbol{\Xi}_{\mathrm R}:\boldsymbol{\Phi}_{\mathrm R}^{H}\boldsymbol{\Xi}_{\mathrm R}+\boldsymbol{\Xi}_{\mathrm R}^{H}\boldsymbol{\Phi}_{\mathrm R}=\mathbf 0\}$ gives the Riemannian gradient
\begin{equation}
\begin{aligned}
    \mathrm{grad}\,f_{\mathrm R}(\boldsymbol{\Phi}_{\mathrm R})
= & \big(\mathbf M_{\mathrm R}\boldsymbol{\Phi}_{\mathrm R}\mathbf Q_{\mathrm R}+\mathbf B_{\mathrm R}\big)
\\&-\boldsymbol{\Phi}_{\mathrm R}\,\mathrm{sym}\!\Big(\boldsymbol{\Phi}_{\mathrm R}^{H}\big(\mathbf M_{\mathrm R}\boldsymbol{\Phi}_{\mathrm R}\mathbf Q_{\mathrm R}+\mathbf B_{\mathrm R}\big)\Big),
\end{aligned}
\end{equation}
with $\mathrm{sym}(\mathbf X)=\tfrac12(\mathbf X+\mathbf X^{H})$. We adopt the QR (or equivalently, polar) retraction to guarantee that the updated matrix $\boldsymbol{\Phi}_{\mathrm R}$ always remains on the complex Stiefel manifold after each iteration. Specifically, for a given point $\boldsymbol{\Phi}_{\mathrm R}$ on the manifold and a tangent step direction $\boldsymbol{\Xi}_{\mathrm R}\in\mathcal{T}_{\boldsymbol{\Phi}_{\mathrm R}}\mathrm{St}$, the retraction is defined as
\begin{equation}
R_{\boldsymbol{\Phi}_{\mathrm R}}(\boldsymbol{\Xi}_{\mathrm R})
= \mathrm{qf}\!\big(\boldsymbol{\Phi}_{\mathrm R}+\boldsymbol{\Xi}_{\mathrm R}\big),
\end{equation}
where the operator $\mathrm{qf}(\cdot)$ extracts the orthonormal (or unitary) factor from the polar or economy-size QR decomposition of its argument. That is, if we compute the decomposition
\begin{equation}
\boldsymbol{\Phi}_{\mathrm R}+\boldsymbol{\Xi}_{\mathrm R} = \mathbf Q_{\mathrm R}\mathbf R_{\mathrm R},
\end{equation}
where $\mathbf Q_{\mathrm R}^{H}\mathbf Q_{\mathrm R}=\mathbf I$ and $\mathbf R_{\mathrm R}$ is upper triangular (or Hermitian positive-definite in the case of a polar decomposition), then the retracted point is given by
\begin{equation}
R_{\boldsymbol{\Phi}_{\mathrm R}}(\boldsymbol{\Xi}_{\mathrm R}) = \mathbf Q_{\mathrm R}.
\end{equation}
This operation projects $\boldsymbol{\Phi}_{\mathrm R}+\boldsymbol{\Xi}_{\mathrm R}$ back onto the Stiefel manifold by finding the closest matrix (in Frobenius norm) with orthonormal columns. In practice, this can be achieved via the standard complex QR decomposition, or through the polar decomposition computed.

The QR/polar retraction is computationally efficient and numerically stable. It preserves orthogonality up to machine precision and is guaranteed to remain within the feasible set defined by $\boldsymbol{\Phi}_{\mathrm R}^{H}\boldsymbol{\Phi}_{\mathrm R}=\mathbf I$. Therefore, at every iteration of the Riemannian optimization process, the updated point $\boldsymbol{\Phi}_{\mathrm R}^{(t+1)} = R_{\boldsymbol{\Phi}_{\mathrm R}^{(t)}}(\alpha_t \boldsymbol{\Xi}_{\mathrm R}^{(t)})$ remains a valid element of the Stiefel manifold. 

\begin{algorithm}[t]
\caption{Optimization of $\boldsymbol{\Phi}_{\mathrm z}$ with $z=R$ or $z=T$ }
\begin{algorithmic}[1]
\State \textbf{Input:} $\{\mathbf w_j\}$, $\{u_{k_r}\}$, $\{t_{k_r}\}$, $\{\mathbf g_{k_r}\}$, $(\mathbf E_{\mathrm z},\mathbf A,\mathbf G,\mathbf N)$; tolerance $\varepsilon$, max iter $T$.
\State Build $\mathbf S_{\mathrm z}$, $\mathbf C_{\mathrm z}$, $\mathbf Q_{\mathrm z}$.
\State For each $k_r$: compute $\mathbf v_{k_r,\mathrm z}$, $\mathbf M_{\mathrm z}$ and $\mathbf B_{\mathrm z}$.
\State Initialize $\boldsymbol{\Phi}_{\mathrm z}^{(0)}$
\For{$t=0,1,\ldots,T-1$}
  \State $\mathbf G_{\mathrm z}=\mathbf M_{\mathrm z}\boldsymbol{\Phi}_{\mathrm z}^{(t)}\mathbf Q_{\mathrm z}+\mathbf B_{\mathrm z}$ \hfill
  \State $\mathbf H_{\mathrm z}=\mathbf G_{\mathrm z}-\boldsymbol{\Phi}_{\mathrm z}^{(t)}\mathrm{sym}\!\big((\boldsymbol{\Phi}_{\mathrm z}^{(t)})^{H}\mathbf G_{\mathrm z}\big)$ \hfill 
  \If{$\|\mathbf H_{\mathrm z}\|_F\le \varepsilon$} \textbf{break} \EndIf
  \State Set search direction $\boldsymbol{\Xi}_{\mathrm z}=-\mathbf H_{\mathrm z}$; select step $\alpha_t$ by backtracking Armijo.
  \State $\mathbf Z=\boldsymbol{\Phi}_{\mathrm z}^{(t)}+\alpha_t\boldsymbol{\Xi}_{\mathrm z}$; compute economy QR/polar: $\mathbf Z=\mathbf Q\mathbf z$.
  \State $\boldsymbol{\Phi}_{\mathrm z}^{(t+1)}\leftarrow \mathbf Q$.
\EndFor
\State \textbf{Output:} $\boldsymbol{\Phi}_{\mathrm z}$.
\end{algorithmic}
\end{algorithm}

\subsubsection{Optimization of $\boldsymbol{\Phi}_T$}

Following a similar approach to the reflective case, we now optimize the transmissive response while fixing $\boldsymbol{\Phi}_{\mathrm{R}}$. The optimization sub-problem becomes:
\begin{subequations}
\begin{equation}
\min_{\boldsymbol{\Phi}_{\mathrm{T}}}
\ \sum_{k_t\in\mathcal{K}_{\mathrm{T}}}
t_{k_t}\,\varepsilon_{k_t}(\boldsymbol{\Phi}_{\mathrm{R}},\boldsymbol{\Phi}_{\mathrm{T}})
\end{equation}
\begin{equation}
\text{s.t.}\quad
\boldsymbol{\Phi}_{\mathrm{T}}^{H}\boldsymbol{\Phi}_{\mathrm{T}}=\mathbf{I},
\end{equation}
\end{subequations}
which constrains $\boldsymbol{\Phi}_{\mathrm{T}}$ to the complex Stiefel manifold. Given $\mathbf{W}=[\mathbf{w}_1,\ldots,\mathbf{w}_{K}]$ and define the fixed matrices:
\begin{subequations}
    \begin{equation}
        \mathbf{S}_{\mathrm{T}}\triangleq \mathbf{E}_{\mathrm{T}}\mathbf{A}\mathbf{G}\mathbf{W}, \quad
        \mathbf{C}_{\mathrm{T}}\triangleq \mathbf{E}_{\mathrm{T}}\mathbf{A}\mathbf{N}\mathbf{A}\mathbf{E}_{\mathrm{T}}^{H} ,
    \end{equation}
    \begin{equation}
        \mathbf{Q}_{\mathrm{T}}\triangleq \mathbf{S}_{\mathrm{T}}\mathbf{S}_{\mathrm{T}}^{H}+\mathbf{C}_{\mathrm{T}}.
    \end{equation}
\end{subequations}
For each $k_t\in\mathcal{K}_{\mathrm{T}}$, let $d_{k_t,j}\triangleq\mathbf{h}_{k_t}^{H}\mathbf{w}_j$, let $\mathbf{s}_{\mathrm{T},j}$ denote the $j$-th column of $\mathbf{S}_{\mathrm{T}}$, and define:
\begin{equation}
\mathbf{v}_{k_t,\mathrm{T}}\triangleq |u_{k_t}|^{2}\sum_{j\in\mathcal{K}}\mathbf{s}_{\mathrm{T},j}\,d_{k_t,j}^{*}\;-\;u_{k_t}\,\mathbf{s}_{\mathrm{T},k_t}\in\mathbb{C}^{n_{\mathrm{T}}}.
\end{equation}
Discarding terms independent of $\boldsymbol{\Phi}_{\mathrm{T}}$, the objective can be written as:
\begin{equation}
    \begin{aligned}
      f_{\mathrm{T}}(\boldsymbol{\Phi}_{\mathrm{T}})
=&\sum_{k_t\in\mathcal{K}_{\mathrm{T}}} t_{k_t}\!\Big(
|u_{k_t}|^{2}\,\mathrm{tr}\!\big(\boldsymbol{\Phi}_{\mathrm{T}}^{H}\mathbf{g}_{k_t}\mathbf{g}_{k_t}^{H}\boldsymbol{\Phi}_{\mathrm{T}}\mathbf{Q}_{\mathrm{T}}\big)
\\&+ 2\,\Re\!\{\mathrm{tr}(\mathbf{v}_{k_t,\mathrm{T}}\mathbf{g}_{k_t}^{H}\boldsymbol{\Phi}_{\mathrm{T}})\}
\Big).  
    \end{aligned}
\end{equation}
The Euclidean gradient is:
\begin{equation}
\nabla f_{\mathrm{T}}(\boldsymbol{\Phi}_{\mathrm{T}})
=\sum_{k_t} t_{k_t}\!\left(
|u_{k_t}|^{2}\,\mathbf{g}_{k_t}\mathbf{g}_{k_t}^{H}\,\boldsymbol{\Phi}_{\mathrm{T}}\,\mathbf{Q}_{\mathrm{T}}
+ 2\,\mathbf{g}_{k_t}\mathbf{v}_{k_t,\mathrm{T}}^{H}
\right).
\end{equation}
Aggregating constants yields:
\begin{subequations}
    \begin{equation}
        \mathbf{M}_{\mathrm{T}}\triangleq\sum_{k_t} t_{k_t}|u_{k_t}|^{2}\,\mathbf{g}_{k_t}\mathbf{g}_{k_t}^{H},\qquad
\mathbf{B}_{\mathrm{T}}\triangleq\sum_{k_t} 2t_{k_t}\,\mathbf{g}_{k_t}\mathbf{v}_{k_t,\mathrm{T}}^{H},
    \end{equation}
    \begin{equation}
        \nabla f_{\mathrm{T}}(\boldsymbol{\Phi}_{\mathrm{T}})=\mathbf{M}_{\mathrm{T}}\boldsymbol{\Phi}_{\mathrm{T}}\mathbf{Q}_{\mathrm{T}}+\mathbf{B}_{\mathrm{T}}.
    \end{equation}
\end{subequations}
Projecting onto the tangent space $\mathcal{T}_{\boldsymbol{\Phi}_{\mathrm{T}}}\mathrm{St}=\{\boldsymbol{\Xi}_{\mathrm{T}}:\boldsymbol{\Phi}_{\mathrm{T}}^{H}\boldsymbol{\Xi}_{\mathrm{T}}+\boldsymbol{\Xi}_{\mathrm{T}}^{H}\boldsymbol{\Phi}_{\mathrm{T}}=\mathbf{0}\}$ gives the Riemannian gradient:
\begin{equation}
\begin{aligned}
    \mathrm{grad}\,f_{\mathrm{T}}(\boldsymbol{\Phi}_{\mathrm{T}})
= & \big(\mathbf{M}_{\mathrm{T}}\boldsymbol{\Phi}_{\mathrm{T}}\mathbf{Q}_{\mathrm{T}}+\mathbf{B}_{\mathrm{T}}\big)
\\&-\boldsymbol{\Phi}_{\mathrm{T}}\,\mathrm{sym}\!\Big(\boldsymbol{\Phi}_{\mathrm{T}}^{H}\big(\mathbf{M}_{\mathrm{T}}\boldsymbol{\Phi}_{\mathrm{T}}\mathbf{Q}_{\mathrm{T}}+\mathbf{B}_{\mathrm{T}}\big)\Big),
\end{aligned}
\end{equation}
with $\mathrm{sym}(\mathbf{X})=\tfrac12(\mathbf{X}+\mathbf{X}^{H})$. We adopt the same QR (or polar) retraction to maintain the Stiefel manifold constraint. For a given point $\boldsymbol{\Phi}_{\mathrm{T}}$ and tangent step direction $\boldsymbol{\Xi}_{\mathrm{T}}\in\mathcal{T}_{\boldsymbol{\Phi}_{\mathrm{T}}}\mathrm{St}$, the retraction is:
\begin{equation}
R_{\boldsymbol{\Phi}_{\mathrm{T}}}(\boldsymbol{\Xi}_{\mathrm{T}})
= \mathrm{qf}\!\big(\boldsymbol{\Phi}_{\mathrm{T}}+\boldsymbol{\Xi}_{\mathrm{T}}\big),
\end{equation}
where $\mathrm{qf}(\cdot)$ extracts the orthonormal factor from the economy-size QR decomposition. That is, if:
\begin{equation}
\boldsymbol{\Phi}_{\mathrm{T}}+\boldsymbol{\Xi}_{\mathrm{T}} = \mathbf{Q}_{\mathrm{T}}\mathbf{R}_{\mathrm{T}},
\end{equation}
where $\mathbf{Q}_{\mathrm{T}}$ satisfies $\mathbf{Q}_{\mathrm{T}}^{H}\mathbf{Q}_{\mathrm{T}}=\mathbf{I}$ and $\mathbf{R}_{\mathrm{T}}$ is an upper triangular, then the retracted point is:
\begin{equation}
R_{\boldsymbol{\Phi}_{\mathrm{T}}}(\boldsymbol{\Xi}_{\mathrm{T}}) = \mathbf{Q}_{\mathrm{T}}.
\end{equation}
The complete optimization procedure to optimze the response of the matrix $\mathbf{\Phi}_z$, where $z = T$ or $z =R$ is given in Algorithm $2$.

\begin{figure}
    \centering
    \includegraphics[width=0.95\linewidth]{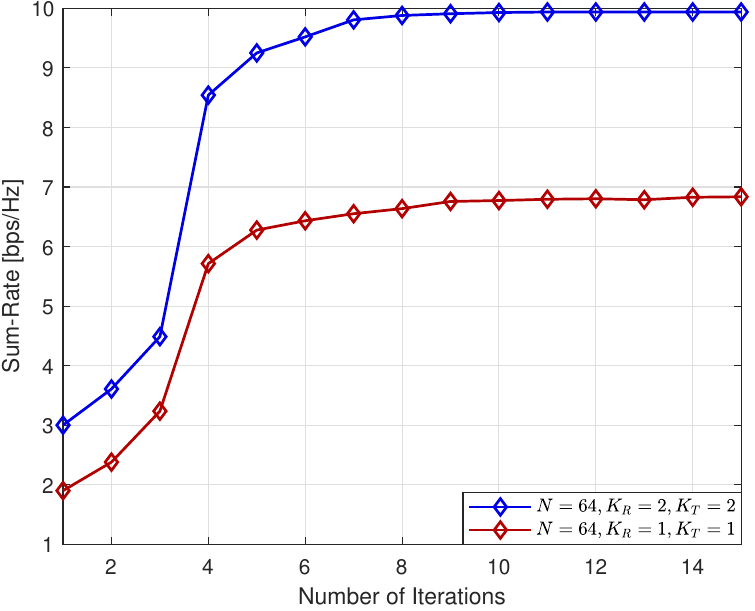}
    \caption{Convergence of the proposed methods at $P_{BS}=20$dBm.}
    \label{fig:convergenza}
\end{figure}

Finally, the overall algorithm to solve the joint problem is formally stated in Algorithm 3. Figure \eqref{fig:convergenza} shows the typical convergence behaviour of the proposed method at transmit power $20$dBm with $2$ or $1$ user per zone.

\begin{algorithm}[t]
\caption{WMMSE–AO for Active STAR BD-RIS Sum-Rate Maximization}
\label{alg:master}
\begin{algorithmic}[1]
\State \textbf{Input:} $\{\mathbf h_k,\mathbf g_k\}$, $\mathbf G$, $\mathbf N$, $P_{\mathrm{BS}}$; init $\{\mathbf w_k^{(0)}\}$, $\boldsymbol{\Phi}_{\mathrm R}^{(0)}$, $\boldsymbol{\Phi}_{\mathrm T}^{(0)}$, $\mathbf A^{(0)}$, $(\mathbf E_{\mathrm R}^{(0)},\mathbf E_{\mathrm T}^{(0)})$; tol $\varepsilon$, max iters $T$.
\For{$t=0,1,\ldots,T-1$}
  \State \textbf{Update Equalizers}: For all $k\in\mathcal K$: compute $u_k^{(t+1)}$ by \eqref{eq:u_mmse_kr_final_corrected}--\eqref{eq:u_mmse_kt_final_corrected}
    \State \textbf{Update MSE weights}: Set $t_k^{(t+1)}=1/\varepsilon_k^{\mathrm{MMSE}}$ by \eqref{eq:t_mmse_kr_final_corrected}.
  \State \textbf{Update Beamformers}: Build $\mathbf Q$ via \eqref{eq:Q_R_corrected}--\eqref{eq:Q_T_corrected}; find $\lambda\ge0$ s.t. BS power holds; update $\{\mathbf w_k^{(t+1)}\}$ by \eqref{eq:wkr_optimal_corrected}--\eqref{eq:wkt_optimal_corrected}.
  \State \textbf{Updated Amplification Matrix:} Solve the convex problem \eqref{prob:wmmse_A_correct} to obtain $\mathbf A^{(t+1)}$.
  \State \textbf{Update Energy splitting:} Call \textbf{Alg.~\ref{alg:energy_splitting}} with current variables to get $(\mathbf E_{\mathrm R}^{(t+1)},\mathbf E_{\mathrm T}^{(t+1)})$.
  \State \textbf{Update BD-RIS responses:} 
  \Statex \quad \textit{(a)} Fix $\boldsymbol{\Phi}_{\mathrm T}^{(t)}$, call \textbf{Alg.~2} with $z{=}R$ to get $\boldsymbol{\Phi}_{\mathrm R}^{(t+1)}$.
  \Statex \quad \textit{(b)} Fix $\boldsymbol{\Phi}_{\mathrm R}^{(t+1)}$, call \textbf{Alg.~2} with $z{=}T$ to get $\boldsymbol{\Phi}_{\mathrm T}^{(t+1)}$.
  \State \textbf{Check Convergence} 
\EndFor
\State \textbf{Output:} $\{\mathbf w_k\},\,\mathbf A,\,(\mathbf E_{\mathrm R},\mathbf E_{\mathrm T}),\,(\boldsymbol{\Phi}_{\mathrm R},\boldsymbol{\Phi}_{\mathrm T})$.
\end{algorithmic}
\end{algorithm}

\subsection{Convergence Analysis}
The proposed algorithm follows an alternating optimization (AO) structure under the WMMSE reformulation in \eqref{prob:wmmse_equiv_formulation_concise}. At each outer iteration, the receiver equalizers and weights are updated in closed form to their MMSE-optimal values, which minimizes the corresponding block exactly and thus cannot increase the global objective. Given these updates, the digital beamforming block is convex and is solved to optimality via the KKT system in \eqref{eq:wkr_optimal_corrected}–\eqref{eq:wkt_optimal_corrected} together with a one-dimensional bisection on the dual variable to satisfy the BS power constraint, again yielding a nonincreasing objective. The amplification matrix subproblem \eqref{prob:wmmse_A_correct} is a convex quadratic program with linear/convex power constraints; solving it by a standard convex solver returns a global minimizer and therefore further decreases (or maintains) the cost. For the energy-splitting variables, the cyclic coordinate-descent procedure evaluates a finite candidate set per coordinate and accepts only updates that strictly reduce the global WMMSE objective; if none is found, the current value is retained. This global acceptance rule guarantees monotonic descent while permitting interior steps that avoid boundary trapping. Finally, the reflective and transmissive BD-RIS response matrices are updated on the complex Stiefel manifold using a Riemannian projected-gradient step with Armijo backtracking and QR/polar retraction. Each accepted step satisfies a sufficient-decrease condition and preserves feasibility, so these blocks are also nonexpansive in the objective. 

\begin{figure*}[t]
    \centering
 \begin{minipage}{0.48\textwidth}
       \centering
    \includegraphics[width=0.9\linewidth]{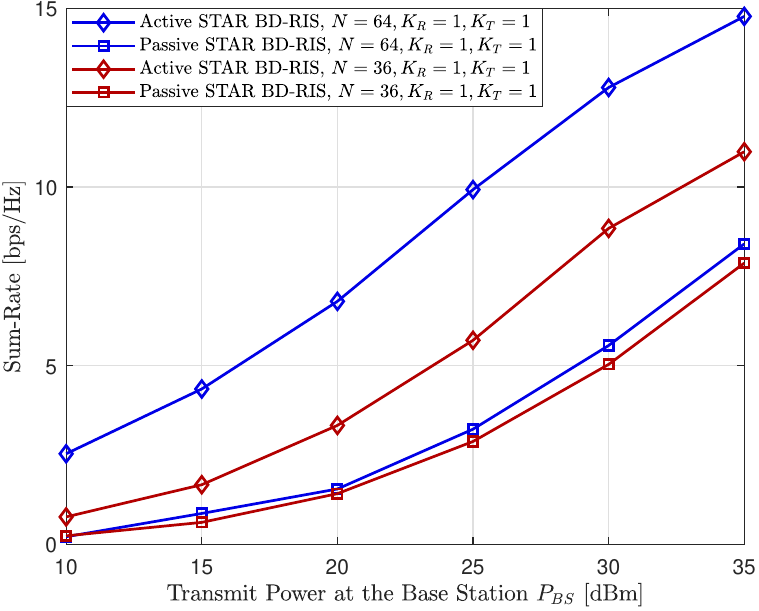}
    \caption{Sum-rate as a function of $P_{\mathrm{BS}}$  with $1$ transmissive and $1$ reflective user.}
    \label{figura_potenza_1_users}
\end{minipage}  
      \begin{minipage}{0.48\textwidth}
       \centering
    \includegraphics[width=0.9\linewidth]{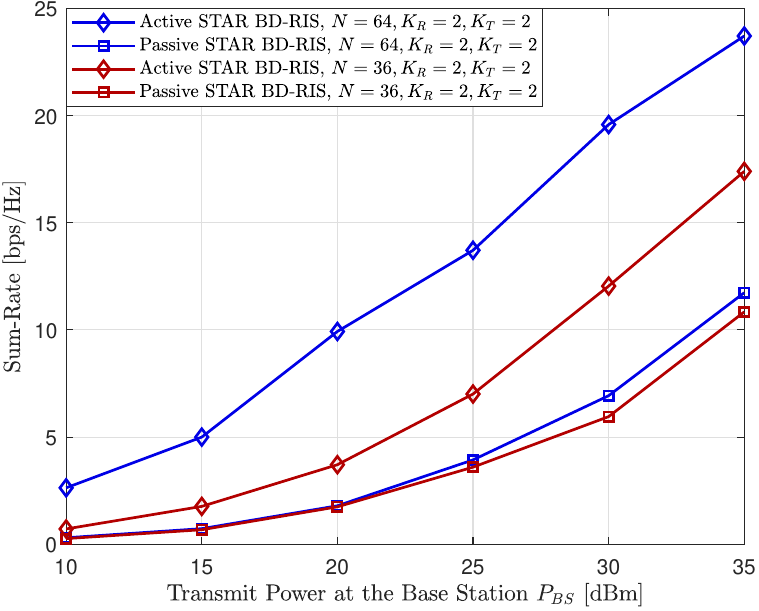}
    \caption{Sum-rate as a function of $P_{\mathrm{BS}}$  with $1$ transmissive and $1$ reflective users.}
    \label{figura_potenza_2_users}
    \end{minipage}  
    \vspace{-4mm}
\end{figure*} 

Since every block update is either globally optimal (convex blocks) or ensures sufficient decrease (manifold and coordinate blocks), the overall WMMSE objective is monotonically nonincreasing across outer iterations. The objective is bounded below because each term $t_k \varepsilon_k - \ln t_k$ is nonnegative at the optimal $t_k$, hence the sequence of objective values converges. Standard AO and Riemannian optimization arguments then imply that every accumulation point of the iterate sequence satisfies blockwise first-order optimality (i.e., the KKT conditions of each subproblem with the remaining blocks fixed). Owing to the exact WMMSE–rate equivalence, such block-stationary points correspond to first-order stationary solutions of the original sum-rate maximization problem \eqref{prob:SR_BDRIS}. In practice, we terminate when the relative decrease in the WMMSE objective falls below a small threshold, which empirically coincides with near-stationary behavior of the primal variables.

 \subsection{Complexity Analysis}
The computational complexity of each outer iteration of the proposed WMMSE–AO algorithm is summarized as follows. The updates of the MMSE equalizers and weights require $\mathcal{O}(K^2 M + K N^2)$ operations, while the digital beamforming step involves constructing and factorizing the curvature matrix $\mathbf{Q}$ at a cost of $\mathcal{O}(M^3)$. The amplification matrix update, with $\mathbf{A}$ being diagonal, leads to a convex quadratic program in $N$ variables whose solution scales as $\widetilde{\mathcal{O}}(N^2)$ due to the diagonal structure. The energy-splitting optimization via cyclic coordinate descent has complexity $\mathcal{O}(N \bar{S} K)$, where $\bar{S}$ is the number of candidate values per element, and the Riemannian updates of $\boldsymbol{\Phi}_{\mathrm{R}}$ and $\boldsymbol{\Phi}_{\mathrm{T}}$ each require $\mathcal{O}(I_{\Phi} N^3)$ operations, executable in parallel. Hence, the total per-iteration complexity is $\mathcal{O}(M^3 + N^2 + I_{\Phi}N^3 + K^2 M + K N^2)$, which is dominated by the active STAR BD-RIS optimization.

\section{Simulation Results} \label{section_4}
In this section, we provide extensive simulation results
to demonstrate the advantages of the proposed multifunctional BD-RIS architecture and illustrate the effectiveness of our proposed algorithm. We assume that the noise level at the RIS and at the users is the same as $\sigma_k = \sigma_i= -90 \text{dBm}, \forall k \in \mathcal{K} $ and $1 \leq i\leq N$. The system is assumed to operate at the carrier frequency $f=3.5$ GHz. The BS transmit antenna array is assumed to be a uniform linear array with antenna
spacing given by $\lambda/2$, where $\lambda$ denotes the wavelength. The channels are modelled with the Rician fading channel model, i.e. $\bmh = \sqrt{\frac{\kappa}{1 + \kappa}} \bmh_{LoS} + \sqrt{\frac{1}{1 + \kappa}} \bmh_{NLoS}$, where $\bmh_{LoS}$, $\bmh_{NLoS}$, and $\kappa$, denote the line-of-sight (LoS), non-LoS, and Rician factor, respectively. For the direct channel $\bmh_k, \forall k$, the path loss exponent is set to $2.9$, and for the channels from BS to users via BD-RIS the pathloss exponent is set to $2$ and $\kappa=3$ dB. The 3D coordinates for the BS and BD-RIS position are chosen to be $p_{BS} = (0, 0, 10)$ and $
p_{RIS} = (180, 0, 5)$, respectively, the users' positions in the reflective and transmissive zones are assumed to be randomly distributed on a circle of radius $40$m centered in RIS's position. The maximum transmit power at the BD-RIS is fixed to be $P_{\max}=10$dBm, and the per-amplifier limit is set the $P_{\max,i} = P_{\max}/N, \forall i$. The number of transmit antennas is set to $10$, and varying BD-RIS sizes are considered. In the following, we label our proposed method as $\emph{Active STAR BD-RIS}$, and as a benchmark, we compare our method with the conventional approach, which deploys \emph{Passive STAR BD-RIS}.

Figure \ref{figura_potenza_1_users} shows the performance for the case of $K_R = 1$, $K_T = 1$, and it is visible that the proposed method consistently outperforms the passive counterpart across all RIS sizes at transmit power levels of the base station. For $N = 64$, at low transmit power ($P_{\text{BS}} = 10$ dBm), the active method achieves approximately $\sim 1100\%$ higher sum-rate than the passive design, while at high power ($P_{\text{BS}} = 35$ dBm), the gain moderates to around $\sim 75.9\%$. Similarly, for $N = 36$, the active approach shows substantial gains of approximately $\sim 225\%$ at low power and $\sim 40\%$ at high power, highlighting its efficacy even with fewer elements. Figure \ref{figura_potenza_2_users} shows the performance for the case of $K_R = 2$, $K_T = 2$. For $N = 64$, the proposed method delivers a sum rate of approximately $\sim 770\%$ higher at low power ($P_{\text{BS}} = 10$ dBm) and $\sim 102\%$ at high power ($P_{\text{BS}} = 35$ dBm), while for $N = 36$, the gains are approximately $\sim 170\%$ at low power and $\sim 60\%$ at high power. These results underscore the significant performance improvements offered by active BD-RIS, particularly in low-power regimes and across varying RIS scales.

\begin{figure*}[t]
    \centering
 \begin{minipage}{0.48\textwidth}
        \centering
    \includegraphics[width=0.9\linewidth]{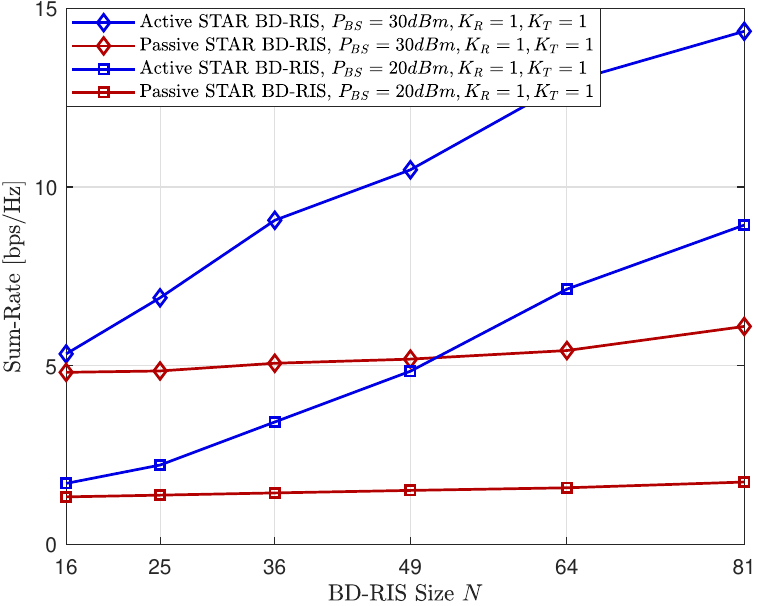}
    \caption{Sum-rate as a function of BD-RIS size with $1$ user per zone.}
    \label{fig:N_2UE}
\end{minipage}  
      \begin{minipage}{0.48\textwidth}
         \centering
    \includegraphics[width=0.9\linewidth]{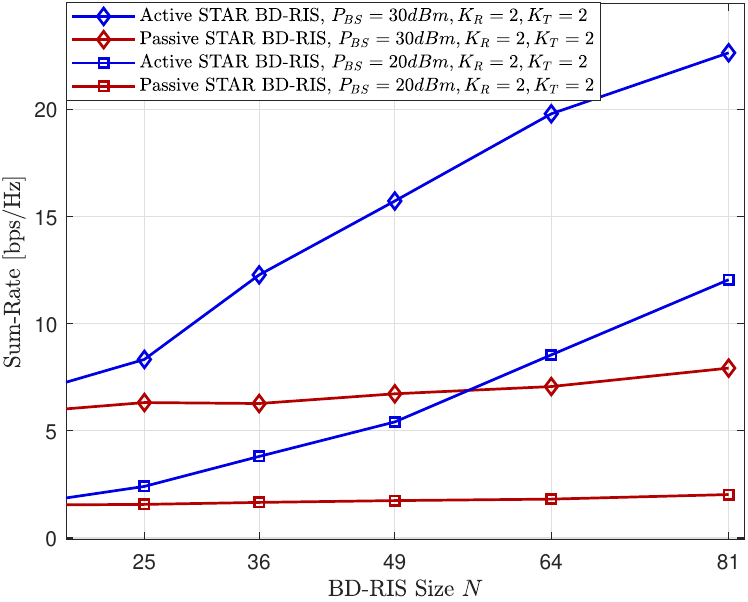}
    \caption{Sum-rate as a function of BD-RIS size with $2$ users per zone.}
    \label{fig:N_4UE}
    \end{minipage}  
    \vspace{-4mm}
\end{figure*} 

 Figure \ref{fig:N_2UE} illustrates the performance of the sum rate versus BD-RIS size \(N\) for the case of two users with \(K_R = 1, K_T = 1\). It is clearly shown that the proposed method consistently outperforms the passive benchmark across all RIS sizes at both considered transmit power levels. For a small RIS size of $N=16$, the proposed method achieves only $\sim 10\%$ and $\sim 28\%$ additional gain at the transmit power $30$dBm and $20$dBm, respectively. However, as the size of the BD-RIS increases, the performance gains increase considerably. Namely, at the BD-RIS of size $81$, the proposed method achieves $\sim 135\%$ and $\sim 412\%$ additional gain at the transmit power levels $30$ dBm and $20$ dBm, respectively. The case of $4$ users with \(K_R = 2, K_T = 2\) is shown in Figure \ref{fig:N_4UE}. We can see that the proposed method for a smaller RIS size $N=16$ achieves $\sim 18\%$ and $\sim 10\%$ additional gains in transmission powers $30$ dBm and $20$ dBm, respectively.  A significant increase is observed as the size of the BD-RIS increases, with the proposed method achieving $\sim 185\%$ and $\sim 340\%$ additional gains in the transmission powers $30$dBm and $20$dBm, respectively. Note that in both figures, the performance of passive BD-RIS varies very slowly, highlighting the potential of the proposed method in all sizes of BD-RIS.
 
\begin{figure}
    \centering
    \includegraphics[width=0.9\linewidth]{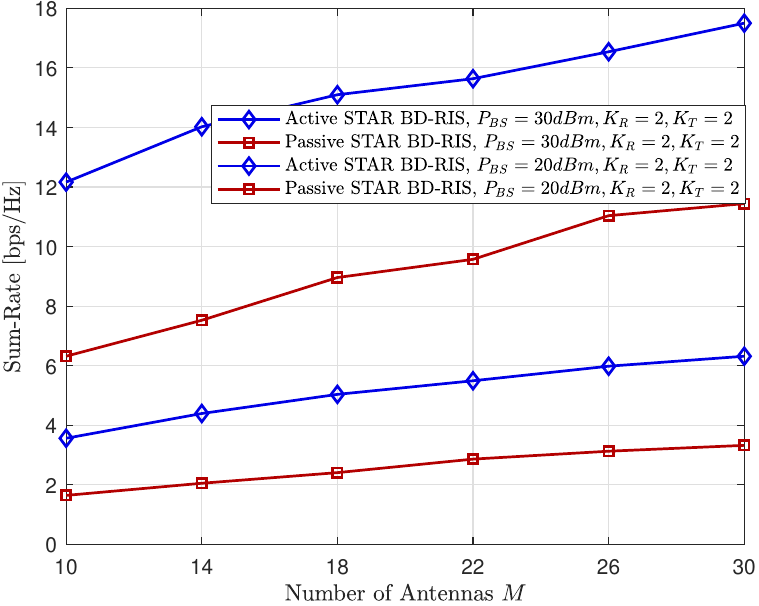}
    \caption{Sum-rate as a function of the number of antennas with BD-RIS size $N=36$.}
    \label{fig:funzione_M} \vspace{-1mm}
\end{figure}

In the following analysis, we fix the BD-RIS size to $M=36$ and evaluate the performance of active and passive BD-RIS architectures as a function of the number of transmit antennas from the base station, as illustrated in Figure \ref{fig:funzione_M}. The results demonstrate that while increasing the number of transmit antennas from $M=10$ to $M=30$ significantly enhances the sum rate for both schemes, the relative performance gap between active and passive BD-RIS remains consistently around $\sim 93\%$ to $\sim 116\%$ across different antenna configurations and power levels. This consistent gain suggests that increasing the BS antenna count benefits both architectures similarly, without changing their relative performance. Hence, achieving further improvements beyond passive BD-RIS is more effectively realized by enlarging the number of active BD-RIS elements rather than merely expanding the BS antenna array.

Overall, the simulation results clearly demonstrate the substantial performance advantages of the proposed architecture over the conventional passive counterpart. The active design consistently achieves higher sum-rate across different transmit power levels, user configurations, and RIS sizes. In particular, the gains are most pronounced in the low-power regime, where active amplification effectively compensates for the inherent signal attenuation of passive structures, yielding improvements exceeding an order of magnitude in some cases. Moreover, as the BD-RIS size increases, the performance gap widens significantly, confirming the scalability and effectiveness of the active architecture. The results with varying base-station antenna numbers further indicate that, while both systems benefit from larger antenna arrays, the relative advantage of the active approach remains stable, underscoring that enhancing the BD-RIS with active elements provides a far more impactful performance boost than merely increasing transmit antennas. Collectively, these findings validate the superiority of the proposed active design in achieving higher spectral efficiency and improved coverage for next-generation wireless systems.

\section{Conclusions} \label{section_5}
This paper proposed a novel STAR BD-RIS architecture with per-element amplification and lossless power splitting, supported by a physically consistent signal model under practical hardware and power constraints. A WMMSE-based alternating optimization algorithm with provable monotonic convergence was developed, featuring closed-form and low-complexity updates for all variable blocks. Simulation results demonstrated substantial sum-rate gains over conventional passive STAR BD-RIS, especially in low-power regimes and for larger RIS sizes, while maintaining stable performance advantages across varying base-station antenna configurations. Overall, the proposed active architecture offers a scalable and efficient means to enhance spectral efficiency, bridging the gap between passive RIS and fully active relaying systems for next-generation wireless networks.

{\footnotesize
\bibliographystyle{IEEEtran}
\bibliography{main}}

\end{document}